\documentclass[%
reprint,
superscriptaddress,
nofootinbib,
amsmath,amssymb,
aps,
]{revtex4-2}

\usepackage{graphicx}
\graphicspath{{figures/}}
\usepackage{dcolumn}
\usepackage{bm}
\usepackage{hyperref}
\usepackage[mathlines]{lineno}
\usepackage[caption=false]{subfig}

\usepackage{color}
\newcommand{\ubc}{Department of Physics \& Astronomy, University of British Columbia, Vancouver, BC V6T 1Z1, Canada}
\newcommand{\meluni}{School of Physics, University of Melbourne, Parkville, VIC 3010, Australia}
\newcommand{\meloz}{OzGrav, University of Melbourne, Parkville, VIC 3010, Australia}
\newcommand{\anu}{OzGrav-ANU, Centre for Gravitational Astrophysics, College of Science, The Australian National University, Australian Capital Territory 2601, Australia}
\newcommand{\bir}{Institute for Gravitational Wave Astronomy \& School of Physics and Astronomy, University of Birmingham, Birmingham, B15 2TT, United Kingdom}

\newcommand{\scF}{$\mathcal{F}$}

\begin{document}


\title{Search for continuous gravitational waves directed at sub-threshold radiometer candidates in O3 LIGO data}

\author{Alan M. Knee}\affiliation{\ubc}
\author{Helen Du}\affiliation{\ubc}
\author{Evan Goetz}\affiliation{\ubc}
\author{Jess McIver}\affiliation{\ubc}
\author{Julian B. Carlin}\affiliation{\meluni}\affiliation{\meloz}
\author{Ling Sun}\affiliation{\anu}
\author{Liam Dunn}\affiliation{\meluni}\affiliation{\meloz}
\author{Lucy Strang}\affiliation{\meluni}
\author{Hannah Middleton}\affiliation{\bir}\affiliation{\meluni}\affiliation{\meloz}
\author{Andrew Melatos}\affiliation{\meluni}\affiliation{\meloz}

\date{\today}

\begin{abstract}
We present results of a follow-up search for continuous gravitational waves (CWs) associated with sub-threshold candidates from the LIGO-Virgo-KAGRA (LVK) All-Sky All-Frequency (ASAF) directed radiometer analysis, using Advanced LIGO data from the third observing run (O3). Each ASAF candidate corresponds to a $1/32\,\,\mathrm{Hz}$ frequency band and ${\sim}13\,\,\mathrm{deg}^2$ sky pixel. Assuming they represent possible CW sources, we analyze all $515$ ASAF candidates using a semi-coherent, \scF-statistic-based matched filter search. The search algorithm incorporates a hidden Markov model (HMM), expanding the signal model to allow frequency spin-wandering, as well as unmodeled frequency evolution of less than $10^{-5}$ Hz per day that is not captured by the searched range of $\pm10^{-9}\,\,\mathrm{Hz/s}$ in frequency derivative. Significance thresholds with a $5\%$ probability of false alarm per ASAF candidate are determined empirically by searching detector noise at various off-target sky positions. We obtain $14$ outliers surviving a set of vetoes designed to eliminate instrumental artifacts. Upon further investigation, these outliers are deemed unlikely to represent astrophysical signals. We estimate the sensitivity of our search to both isolated and binary sources with orbital period greater than one year by recovering simulated signals added to detector data. The minimum detectable strain amplitude at $95\%$ confidence for isolated (long-period binary) sources is $h_0^{95\%} = 8.8\times 10^{-26}$ ($9.4\times 10^{-26}$) at a frequency of $222.6\,\,\mathrm{Hz}$. While this study focuses on ASAF sub-threshold candidates, the method presented could be applied to follow up candidates from future all-sky CW searches, complementing currently existing methods.
\end{abstract}

\maketitle

\section{Introduction} \label{sec:intro}

Continuous gravitational waves (CWs) are an as-yet undiscovered class of persistent, quasi-monochromatic gravitational waves (GWs) \cite{Lasky:2015uia, Sieniawska:2019hmd, Haskell:2021ljd, Riles:2022wwz}. The canonical sources of CWs in the observing band of ground-based GW detectors are rapidly spinning, non-axisymmetric neutron stars, which may be isolated or part of binary systems. Possible emission mechanisms include small deformations of the neutron star crust created by thermoelastic \cite{1971AnPhy..66..816B, 1998ApJ...501L..89B, 2000MNRAS.319..902U, Johnson-McDaniel:2012wbj}, magnetic \cite{PhysRevD.66.084025, 2011MNRAS.417.2288M, 2013PhRvD..88j3005L}, or tectonic \cite{2006ApJ...652.1531M, 2010MNRAS.407L..54C, Giliberti:2021zxn, Kerin:2022ita} stresses, fluid oscillation modes (e.g.,~$r$-modes) that are unstable to gravitational radiation \cite{Owen:1998xg, 1999ApJ...516..307A, 2003ApJ...591.1129A}, and pulsar spin-up glitches \cite{2008CQGra..25v5020V, 2010MNRAS.409.1705B}. A confirmed CW detection would offer novel insight into the physics of ultra-dense matter, allowing for new probes of the nuclear equation-of-state, measurements of neutron star ellipticities, and strong-field tests of gravity. 

Current GW detectors, such as the Advanced Laser Interferometer Gravitational-wave Observatory (LIGO) \cite{LIGOScientific:2014pky}, Advanced Virgo \cite{VIRGO:2014yos}, and KAGRA \cite{KAGRA:2018plz}, may be sensitive enough to detect CW sources within our galaxy. Searches for CW signals in detector data have targeted known young or millisecond pulsars \cite{LIGOScientific:2017hal, LIGOScientific:2017ytx, LIGOScientific:2019xqs, LIGOScientific:2019mhs, LIGOScientific:2020gml, LIGOScientific:2021hvc, LIGOScientific:2021quq, Fesik:2020tvn}, young supernova remnants \cite{LIGOScientific:2018esg, Papa:2020vfz, LIGOScientific:2021mwx, PhysRevD.102.083025}, low-mass X-ray binaries \cite{KAGRA:2022dqk, LIGOScientific:2021ozr, Middleton:2020skz}, and the galactic centre \cite{Piccinni:2019zub, KAGRA:2022osp}. Various all-sky searches have also been carried out, scanning for CW emission from unknown neutron stars in isolated \cite{LIGOScientific:2017wva, LIGOScientific:2019yhl, KAGRA:2021una, KAGRA:2022dwb, Steltner:2023cfk} or binary \cite{LIGOScientific:2020qhb, Covas:2020nwy} configurations. Although there have been no CW detections reported thus far, improvements in detector sensitivity are allowing searches to probe deeper into the physically interesting regions of parameter space.

\begin{figure}
\centering
\includegraphics[width=1\columnwidth]{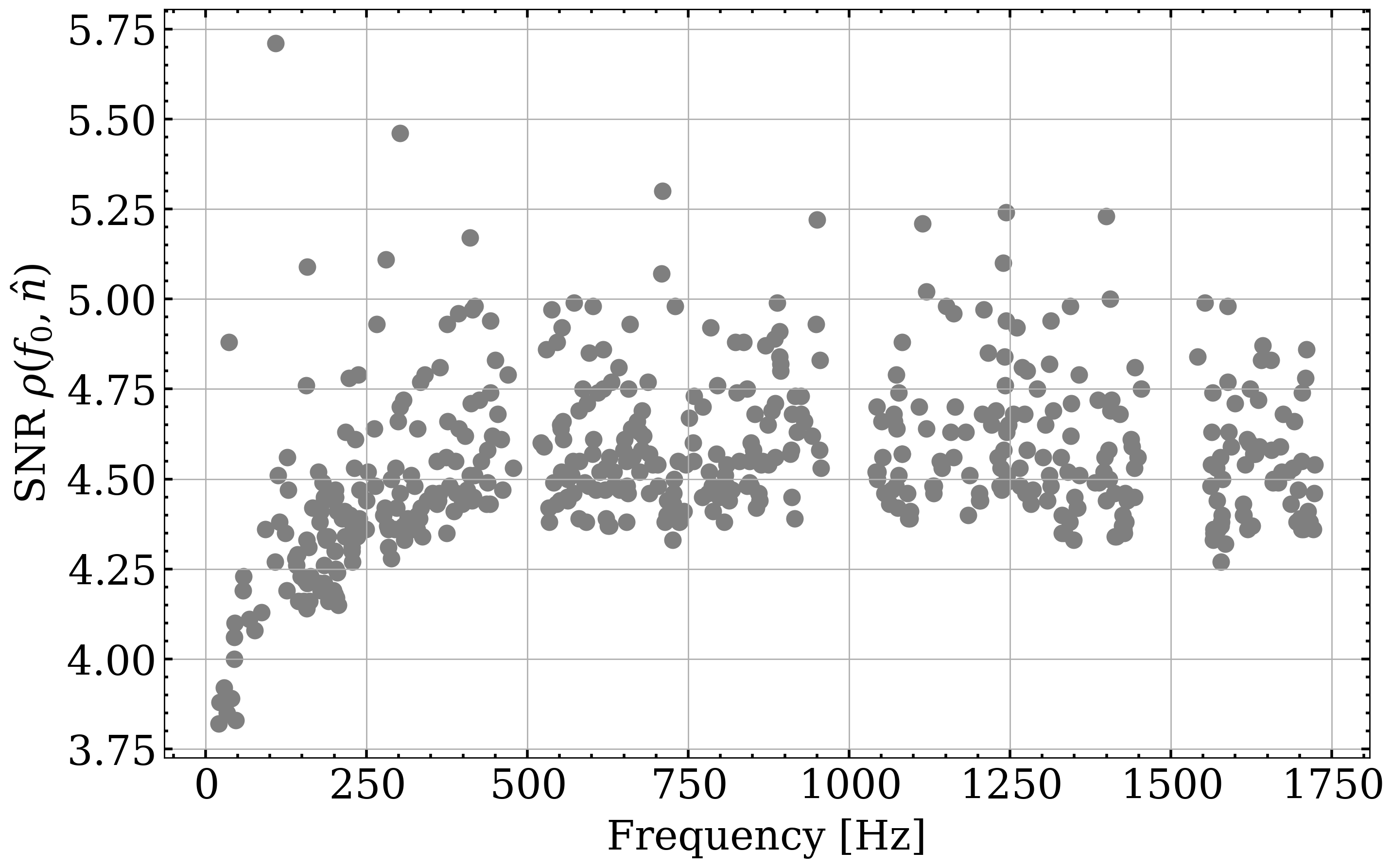}
\caption{\label{fig:candidates} SNR, $\rho(f_0,\boldsymbol{\hat{n}})$, versus central sub-band frequency, $f_0$ of the 515 sub-threshold candidates from the LVK ASAF analysis, which are followed up in this work. The SNR is a function of $f_0$ and sky position, $\boldsymbol{\hat{n}}$, where the latter symbol denotes a unit vector pointing at the candidate sky pixel. The bands around $500$, $1000$, and $1500\,\,\mathrm{Hz}$ are affected by optic suspension harmonics and have been notched out in the ASAF analysis.}
\end{figure}

In this work, we carry out a search for CWs by following up sub-threshold candidates identified by the LIGO-Virgo-KAGRA (LVK) All-Sky All-Frequency (ASAF) radiometer analysis \cite{KAGRA:2021rmt} (hereafter the ``ASAF analysis''). While all-sky CW searches typically assume a deterministic signal coherent over ${\sim}10^3\,\,\mathrm{s}$ at minimum, and extending to even longer coherence times in the later follow-up stages \cite[e.g.,][]{KAGRA:2022dwb}, the ASAF analysis is an unmodeled search with comparatively weaker phase-coherence constraints. In the ASAF method, $192\,\,\mathrm{s}$ data chunks from pairs of detectors are frequency-binned and cross-correlated with sky-dependent phase and amplitude modulations, allowing one to search for a directional, narrowband stochastic signal. Adopting an equal-area tiling of the sky with $3072$ pixels (${\approx}13.43\,\,\mathrm{deg}^2$ per pixel) and a coarse-grained $1/32\,\,\mathrm{Hz}$ frequency resolution, the ASAF analysis searched every frequency sub-band and sky-pixel pair from $20\,\,\mathrm{Hz}$ to $1726\,\,\mathrm{Hz}$ using LIGO-Virgo data from the first three observing runs (with the exception of a few bands excluded due to noise contamination). The ASAF analysis assumed a global signal-to-noise ratio (SNR) threshold with a $5\%$ probability of false alarm over the full band. Although no candidates were found above this threshold, the ASAF analysis identified $515$ sub-threshold candidates (hereafter called ``candidates'') with SNRs exceeding the $99$th SNR percentile averaged over $10\,\,\mathrm{Hz}$ frequency bins \cite{asafcand}. Each candidate thus comprises a frequency sub-band and sky-pixel pairing, making it possible to conduct outlier follow-up with a CW search algorithm. The candidate SNRs and central sub-band frequencies are shown in Fig.~\ref{fig:candidates}. As the ASAF analysis can potentially detect coherent, quasi-monochromatic signals, these candidates represent interesting targets for CW follow-up with more sensitive matched-filtering techniques. If any candidates are CW sources, then a matched-filter search with more demanding coherence requirements will improve the probability of detecting a signal. This is the first time CW analysis techniques have been used to follow up candidates from a stochastic GW search.

Due to the model-agnostic approach of the ASAF analysis, the candidates could exhibit complicated or unpredictable frequency evolution, deviating from the canonical signal model assumed in many CW analyses. Several phenomena can give rise to such behaviour, including stochastic spin-wandering, also known as {\it timing noise} \cite{2010MNRAS.402.1027H, 2010ApJ...725.1607S, 2010ApJ...725.1607S, 2012MNRAS.426.2507P, Parthasarathy:2019txt, Namkham:2019kuw, Lower:2020mjq, Ashton:2014qya}, spin-up glitches \cite{2008CQGra..25v5020V, 2010MNRAS.409.1705B}, and long-period binary motion \cite{Manchester:2004bp}, all of which may degrade the SNR if left unaccounted for, though we leave a quantitative assessment of their impact on CW searches to future work. A signal of this kind could be missed in an all-sky CW search while being marginally detected in a stochastic search, where the assumption of a deterministic phase is relaxed. Motivated by the need for a more flexible signal model, we follow up the $515$ ASAF candidates using a search algorithm which combines the semi-coherent \scF-statistic \cite{Jaranowski:1998qm, Prix:2006wm} with a hidden Markov model (HMM), adapted from the method described in Refs.~\cite{Suvorova:2016rdc, 2018PhRvD..97d3013S} and applied to various CW targets to date \cite[e.g.,][]{KAGRA:2022dqk, LIGOScientific:2021mwx, LIGOScientific:2021ozr, Middleton:2020skz, PhysRevD.102.083025, 2019PhRvD.100b3006B, LIGOScientific:2021rnv, PhysRevD.103.083009, Beniwal:2022nam, Vargas:2022mvs}. The HMM can track a signal as it wanders in frequency over the course of the observing period, broadening the CW signal model to include a wide variety of possible frequency evolution. We carry out this search on publicly available Advanced LIGO data from the third observing run (O3) \cite{KAGRA:2023pio, selfgating, segments, lalsuite}.

This paper is organized as follows: The data set used is described in Sec.~\ref{sec:data}; in Sec.~\ref{sec:method}, we review the implementation of our search algorithm; in Sec.~\ref{sec:sig}, we describe our procedure for identifying CW outliers and screening likely instrumental artifacts; in Sec.~\ref{sec:results}, we present the results of our follow-up searches and estimate the sensitivity; finally, we conclude in Sec.~\ref{sec:conclusion}.

\section{Data set} \label{sec:data}

We perform the search on O3 data \cite{KAGRA:2023pio} from the two Advanced LIGO detectors: LIGO Hanford (H1), and LIGO Livingston (L1). The O3 run began on April 1, 2019, 15:00 UTC (GPS time $1238166018.0$), and ended on March 27, 2020, 17:00 UTC (GPS time $1269363618.0$). The run was split into two parts, O3a and O3b, separated by a month-long commissioning period lasting from October 1 to November 1, 2019, when the detectors were nominally not in observing mode. During O3, the LIGO detectors operated at higher sensitivity and with a higher duty cycle than the previous O1 and O2 runs \cite{2020PhRvD.102f2003B}. As including O1 and O2 data would increase the computational demands of the search without providing a significant enhancement of sensitivity, these data are not used. Similarly, we do not include data from Advanced Virgo or KAGRA due to their lower sensitivities \cite{Virgo:2022ysc, KAGRA:2020agh}.

The search algorithm processes a set of Tukey-windowed, short Fourier transforms (SFTs) from both detectors, taken over 1800 s of calibrated strain data \cite{sfts, selfgating, segments, lalsuite}. We use the C01 calibration version O3 data set \cite{Sun:2020wke, Sun:2021qcg}, with loud noise transients removed via a self-gating procedure \cite{selfgating} and $60$ Hz power harmonics subtracted out \cite{covas2018identification}. Time segments determined to have serious data quality issues (Category 1 data quality vetoes) are not analyzed \cite{davis2021ligo, segments}. Any SFTs generated from segments with significant deadtime due to gating ($30\,\,\mathrm{s}$ or longer) are also not used by our analysis. With these criteria, the total number of SFTs processed by our search is $11025$ for H1 and $10133$ for L1, covering $69.6\%$ and $64.0\%$ of the O3 run, respectively.

\section{Search method} \label{sec:method}

The search algorithm used to follow up each of the ASAF candidates consists of two main analysis steps. Firstly, we filter the O3 data against a set of waveform templates covering the candidate sub-band and sky pixel, using the \scF-statistic as the matched filter statistic. We then use the HMM formalism to reconstruct the optimal frequency path for each search template. We review the \scF-statistic and HMM tracking technique in Sec.~\ref{sec:fstat} and Sec.~\ref{sec:hmm}, respectively. We construct template grids empirically for every candidate, as described in Sec.~\ref{sec:grid}.

\subsection{Semi-coherent \scF-statistic} \label{sec:fstat}

The first step in our analysis is a matched filter search using a frequency domain estimator known as the \scF-statistic. The \scF-statistic is a maximum likelihood statistic for detecting CW emission from spinning neutron stars, modelled as triaxial ellipsoids which rotate about one of their principal axes \cite{Jaranowski:1998qm, Prix:2006wm}. It is denoted by $2\mathcal{F}(x|\boldsymbol{\lambda})$, where $x$ represents the strain data, and $\boldsymbol{\lambda}=\{\alpha,\delta,f,\dot{f}\}$ are the phase parameters of the filter template, consisting of a right ascension, $\alpha$, declination, $\delta$, GW frequency, $f$, and frequency derivative, $\dot{f}\equiv {\rm d}f/{\rm d}t$.  We do not include frequency derivatives beyond $\dot{f}$ within the \scF-statistic, but higher derivatives can be absorbed to some extent into the stochastic variations allowed by the HMM (see Sec.~\ref{sec:hmm}). Modulations of the signal frequency, caused by a non-zero $\dot{f}$ as well as the motion of the Earth relative to the source, are accounted for within the \scF-statistic calculation by demodulating the data with respect to the search template. Our pipeline leverages the \texttt{lalpulsar\_ComputeFstatistic\_v2} implementation of the multi-detector \scF-statistic \cite{cfsv2, lalsuite}, which processes a set of SFTs as input and evaluates the matched filter on a grid of templates. The placement of templates for the ASAF candidates is described in Sec.~\ref{sec:grid}.

We divide the O3 observing period into $N_T=361$ contiguous intervals of duration $T_{\rm coh}=86400\,\,\mathrm{s} = 1\,\,\mathrm{d}$, and evaluate the \scF-statistic coherently over each of these intervals. Phase information is thus preserved within each interval, but is otherwise not tracked between them, i.e.~our search is semi-coherent. The choice of $T_{\rm coh}$ used in our search improves upon the $192\,\,\mathrm{s}$ coherence of the ASAF analysis, achieving a reasonable trade-off between computational cost and sensitivity. Note that there are a total of 32 one-day segments from O3 with no SFTs from either detector, all but one being due to the O3 commissioning break. For days where no data are available, we substitute $2\mathcal{F}=4$ uniformly for all frequencies \cite{LIGOScientific:2019lyx}, consistent with the expectation value of the \scF-statistic in pure Gaussian noise. This procedure allows the HMM, described next in Sec.~\ref{sec:hmm}, to traverse data gaps in an unbiased manner, following previous CW HMM searches \cite[e.g.,][]{LIGOScientific:2021mwx, LIGOScientific:2021ozr}.

\subsection{HMM tracking} \label{sec:hmm}

The next step in our analysis is to find the optimal frequency path for each grid template $(\alpha,\delta,\dot{f})$ by means of HMM tracking \cite{Suvorova:2016rdc, 2018PhRvD..97d3013S}. HMMs are statistical models for inferring the behaviour of an unobserved (``hidden'') state variable, $q(t)$, based on an observed state variable, $o(t)$, under the assumption that the observed state is influenced by the hidden state. Both states take on a sequence of values at $N_T$ discrete epochs, $t_n\in \{t_1,\ldots,t_{N_T}\}$. The hidden state is modeled as a stochastic Markov process, jumping between $N_Q$ discrete values $q(t)\in\{q_1,\ldots,q_{N_Q}\}$ with some transition probability matrix $A_{q_jq_i} = P(q(t_n)=q_j|q(t_{n-1})=q_i)$. At each epoch $t_n$, the observed state $o(t_n)$ is related to $q(t_n)$ through the emission probability, $L_{o(t_n)q_i} = P(o(t_n)|q(t_n)=q_i)$. 

Applied to CW analyses, $q(t)$ represents the heretofore unknown frequency of a CW signal, $f(t)$, which occupies one frequency bin $q_i$ at any given epoch. The observed states, $o(t)$, represent the strain data collected in each of the $N_T$ coherent intervals of a semi-coherent search, where the interval duration is fixed by the coherence time, $T_{\rm coh}=1\,\,\mathrm{d}$, of the matched filter. 
We use the \scF-statistic, conditional on a $(\alpha,\delta,\dot{f})$ template, to map $o(t)$ onto $q(t)$, i.e.~it is the emission probability for the HMM. The spin-wandering model is specified through the transition probability matrix, $A_{q_jq_i}$. As in previous HMM CW searches \cite{Suvorova:2016rdc, PhysRevD.102.083025, LIGOScientific:2021mwx, Middleton:2020skz, LIGOScientific:2021ozr, KAGRA:2022dqk}, we define the transition matrix as
\begin{equation}
    A_{q_jq_i} = \frac{1}{3}(\delta_{q_jq_{i-1}} + \delta_{q_jq_i} + \delta_{q_jq_{i+1}})\,,
\end{equation}
which restricts $q(t)$ to jump by $0$ or $\pm 1$ frequency bins $\Delta f=q_i-q_{i-1}$ over each epoch with equal probability. Since $q(t)$ can jump by no more than one frequency bin per coherent interval, the maximum amount of spin-wandering allowed by the HMM is limited to
\begin{equation}
    |\dot{q}|_{\rm max} = \Delta f/T_{\rm coh}\,.
\end{equation}
The probability that the signal follows some frequency path, $Q=\{q(t_1),\ldots,q(t_{N_T})\}$, conditioned on the observed data, $O=\{o(t_1),\ldots,o(t_{N_T})\}$, is given by
\begin{equation}
    P(Q|O) = \Pi_{q(t_0)}\prod_{n=1}^{N_T} L_{o(t_n)q(t_n)}A_{q(t_n)q(t_{n-1})}\,,
\end{equation}
where $\Pi_{q(t_0)}$ is a prior probability on the initial frequency, $q(t_0)$, taken to be a uniform function of frequency within the search band, $\Pi_{q(t_0)}\propto N_Q^{-1}$.

\begin{figure*}
\centering
\includegraphics[width=2\columnwidth]{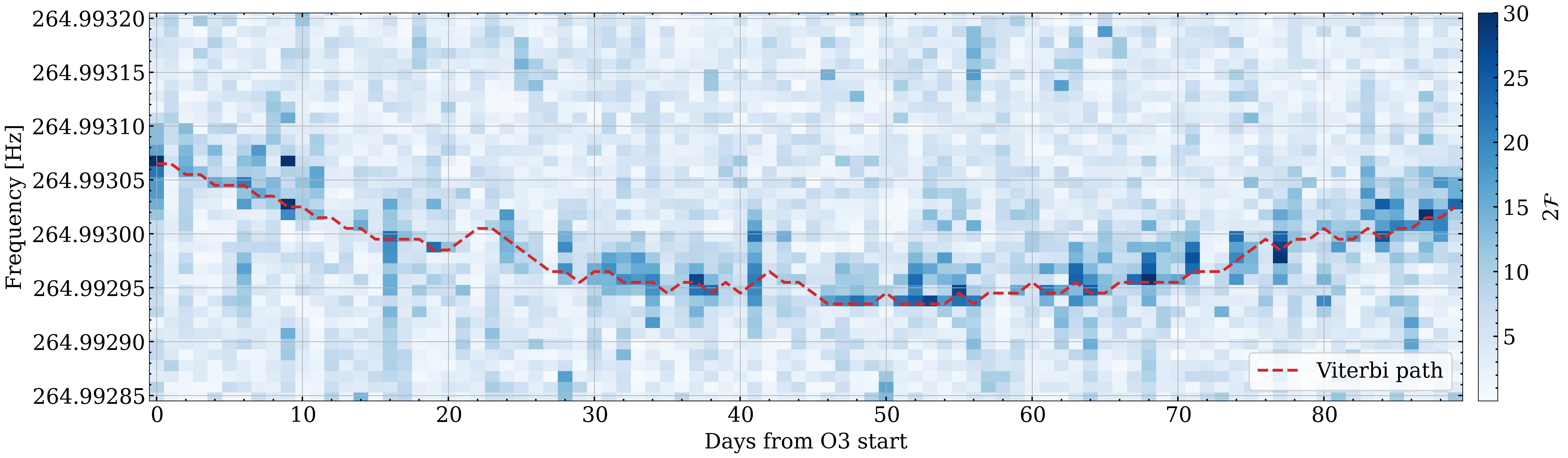}
\caption{\label{fig:viterbi} Recovery of a simulated signal from a long-period binary system injected into three months of Advanced LIGO data from O3, using the search algorithm employed in this work to follow-up the ASAF candidates. The pixel colours indicate the value of the \scF-statistic (i.e.~emission probabilities) in each $T_{\rm coh}=1\,\,\mathrm{d}$ interval (horizontal axis) and $\Delta f=10^{-5}\,\,\mathrm{Hz}$ frequency bin (vertical axis) after demodulating by the template $\dot{f}$ and sky position. The frequency path recovered by the Viterbi algorithm is shown by the red curve. The binary orbit has eccentricity $e=0.4$ with period $P_{\rm b}=10\,\,\mathrm{yr}$. The signal is circularly polarized with strain amplitude $h_0=8\times 10^{-26}$.}
\end{figure*}

The objective of the HMM is to determine the optimal frequency path, $Q=Q^*$, that maximizes $P(Q|O)$. This is accomplished using the Viterbi algorithm---a dynamic programming algorithm which efficiently solves the HMM by strategically discarding sub-optimal paths \cite{1054010}. The detection statistic associated with $Q^*$ (the Viterbi path) is the log-likelihood,
\begin{equation} \label{eq:loglike}
    \mathcal{L}\equiv\log P(Q^*|O)\,,
\end{equation}
which is equivalent to incoherently summing \scF-statistics along the frequency path, up to a linear offset. Although the Viterbi algorithm outputs $N_Q$ optimal paths, each one terminating in a different frequency bin, we maximize over all such paths and report only the globally optimal path for each $(\alpha,\delta,\dot{f})$ template. In Fig.~\ref{fig:viterbi}, we show an example of using the Viterbi algorithm, paired with the \scF-statistic, to recover a simulated signal injected into O3-era detector noise. In this example, the source lies in long-period binary system, resulting in a slow Doppler modulation of the signal frequency which is successfully tracked by the Viterbi algorithm. We refer the reader interested in the full mathematical details of the Viterbi algorithm and HMMs to Ref.~\cite{Suvorova:2016rdc}.

We have chosen to use $T_{\rm coh}=1\,\,\mathrm{d}$, which prevents us from tracking any frequency evolution happening on shorter scales. For reference, the low-mass X-ray binary Scorpius X-1 \cite{Steeghs:2001rx} is thought to exhibit relatively high amounts of spin-wandering, in which $T_{\rm coh}=10\,\,\mathrm{d}$ has been the typical spin-wandering timescale adopted by past Viterbi searches \cite{2017PhRvD..95l2003A, LIGOScientific:2019lyx, KAGRA:2022dqk, Middleton:2020skz, LIGOScientific:2021ozr}. Thus, our Viterbi implementation has enough flexibility to capture the spin-wandering behaviour of most known neutron stars.

\subsection{Template grids} \label{sec:grid}

We search every ASAF candidate using a rectilinear grid of $(\alpha,\delta,\dot{f})$ templates spanning the candidate sub-band and sky pixel. This $1/32\,\,\mathrm{Hz}$ sub-band is further subdivided into smaller frequency bins $\Delta f$. We then evaluate the \scF-statistic for each grid template and $\Delta f$ bin using the semi-coherent algorithm summarized previously in Sec.~\ref{sec:fstat}. We derive fixed grid resolutions empirically for each candidate, as described in Sec.~\ref{sec:freqres} and Sec.~\ref{sec:skyres} for frequency templates and sky templates, respectively. We opt to not use the standard parameter-space metric \cite{Prix:2006wm}, as the spin-wandering allowed by the HMM signal model introduces additional correlations between $\dot{f}$ and sky position templates. We verify that our template grid appropriately covers the required search space for each candidate in Sec.~\ref{sec:mismatches}.

\subsubsection{Frequency resolution} \label{sec:freqres}

We divide each $1/32\,\,\mathrm{Hz}$ candidate sub-band, centered on a frequency $f_0$, into bins of width $\Delta f=10^{-5}\,\,\mathrm{Hz}$. We search over various $\dot{f}$ templates covering a $\pm 10^{-9}\,\,\mathrm{Hz/s}$ range, with a spacing of $\Delta\dot{f}=2\times 10^{-10}\,\,\mathrm{Hz/s}$ between templates. The bounds on the $\dot{f}$ range correspond roughly to the maximum spin-down/up a signal could have and still be contained within one $1/32\,\,\mathrm{Hz}$ ASAF sub-band. By searching over $\dot{f}$, the \scF-statistic demodulation can correct for secular, long-term frequency trends over the observing period, leaving the HMM to track residual drifts occurring on timescales of order $T_{\rm coh}$. Without including $\dot{f}$ in the \scF-statistic, the maximum frequency derivative recoverable by the HMM would be $|\dot{q}|_{\rm max}$, requiring one to either decrease $T_{\rm coh}$ or increase $\Delta f$ to accommodate faster $\dot{f}$, either of which would make the search less sensitive. Combining the HMM with $\dot{f}$ demodulation alleviates this restriction. We found that setting $\Delta\dot{f} \ll 2\Delta f/T_{\rm coh}$ introduced correlations between adjacent $\dot{f}$ templates, and hence chose $\Delta\dot{f}=2\times 10^{-10}\,\,\mathrm{Hz/s}$ to mitigate these correlations. Finally, the $\dot{f}$ values are defined with respect to a reference time, which we take to be the midpoint of O3 (GPS time $1253764818.0$).

\subsubsection{Sky resolution} \label{sec:skyres}

The resolution of the sky grid depends primarily on the frequency and secondarily on the sky position of the ASAF candidate. We calibrate the sky grids by running the search algorithm on simulated signals and then examining the distribution of signal power around the injection point. We first estimate fiducial grid resolutions, $(\Delta\alpha_i^{\rm fid},\Delta\delta_i^{\rm fid})$, across the entire sky for a fixed $1/32\,\,\mathrm{Hz}$ frequency band centered at $1\,\,\mathrm{kHz}$, where $i$ denotes the sky pixel index. We adopt the same \texttt{HEALPix}\footnote{\url{http://healpix.sourceforge.net}.} \cite{2005ApJ...622..759G, Zonca2019} pixel basis as the ASAF analysis, so the pixel indices range from $i\in\{1,\ldots,3072\}$. After selecting a subset of $386$ pixels uniformly spaced across the sky, we simulate a signal at the origin of each sample pixel with phase parameters $(f,\dot{f})=(1\,\,\mathrm{kHz}, 0\,\,\mathrm{Hz/s})$, injecting into synthetic Gaussian noise SFTs created with \texttt{lalpulsar\_Makefakedata\_v5} \cite{lalsuite} for H1 and L1. The generated SFTs include the same time gaps as the real O3 data set, which is done to make the analysis as similar as possible to our O3 search. To ensure the signals can be confidently recovered, we fix $\sqrt{S_{\rm n}}/h_0=1$, where $h_0$ is the signal amplitude and $S_{\rm n}$ is the one-sided power spectral density (PSD) of the detector noise. We search each pixel on a dense sky grid with  $\Delta\alpha=\Delta\delta=0.002\,\,\mathrm{rad}$ using the search algorithm described in Sec.~\ref{sec:fstat}--\ref{sec:hmm}. For these injections, we use the same frequency spacing given in Sec.~\ref{sec:freqres} but do not search over $\dot{f}$, which is fixed at $\dot{f}=0$. We compute the mismatch for each sky template as
\begin{equation} \label{eq:mismatch}
    \mu = 1-\mathcal{L}/\mathcal{L}_0\,,
\end{equation}
where $\mathcal{L}$ is the Viterbi log-likelihood statistic (Eq.~\ref{eq:loglike}) for the template and $\mathcal{L}_0$ is the log-likelihood of a template with identical parameters to the injection. 

\begin{figure}
\centering
\includegraphics[width=1\columnwidth]{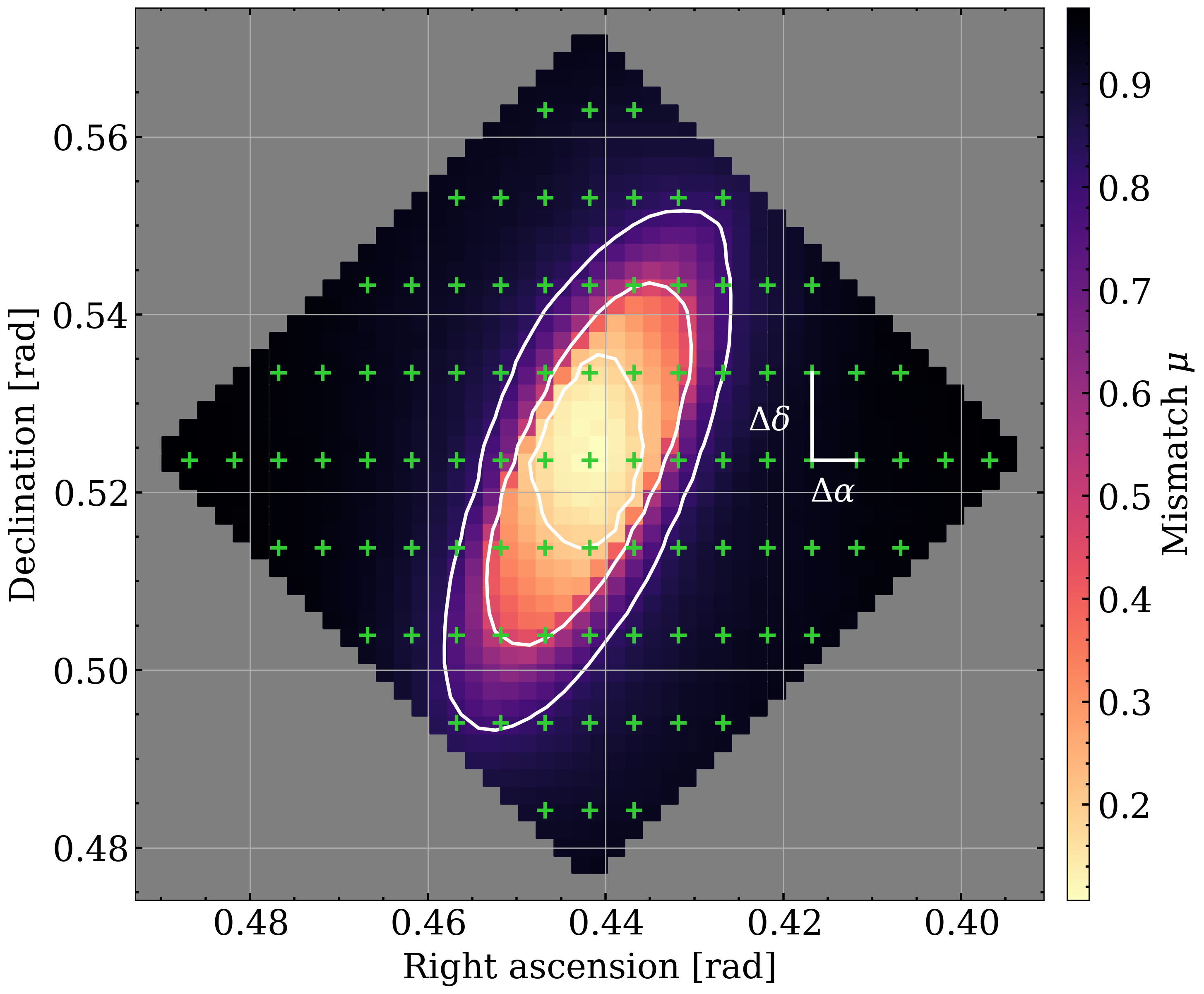}
\caption{\label{fig:skygrid} Example of a fiducial sky grid constructed using a simulated signal at $f_0=1\,\,\mathrm{kHz}$, $\alpha_0=0.442\,\,\mathrm{rad}$, $\delta_0=0.524\,\,\mathrm{rad}$. The diamond-shaped search region is the sky pixel. The mismatch as a function of sky position, $\mu$, is indicated by the colour scale. The $\mu=0.2$, $0.5$, and $0.8$ contours are shown by the white outlines. The green crosses show the resulting sky grid estimated from the $\mu=0.2$ contour. }
\end{figure}

\begin{figure}
\centering
\includegraphics[width=1\columnwidth]{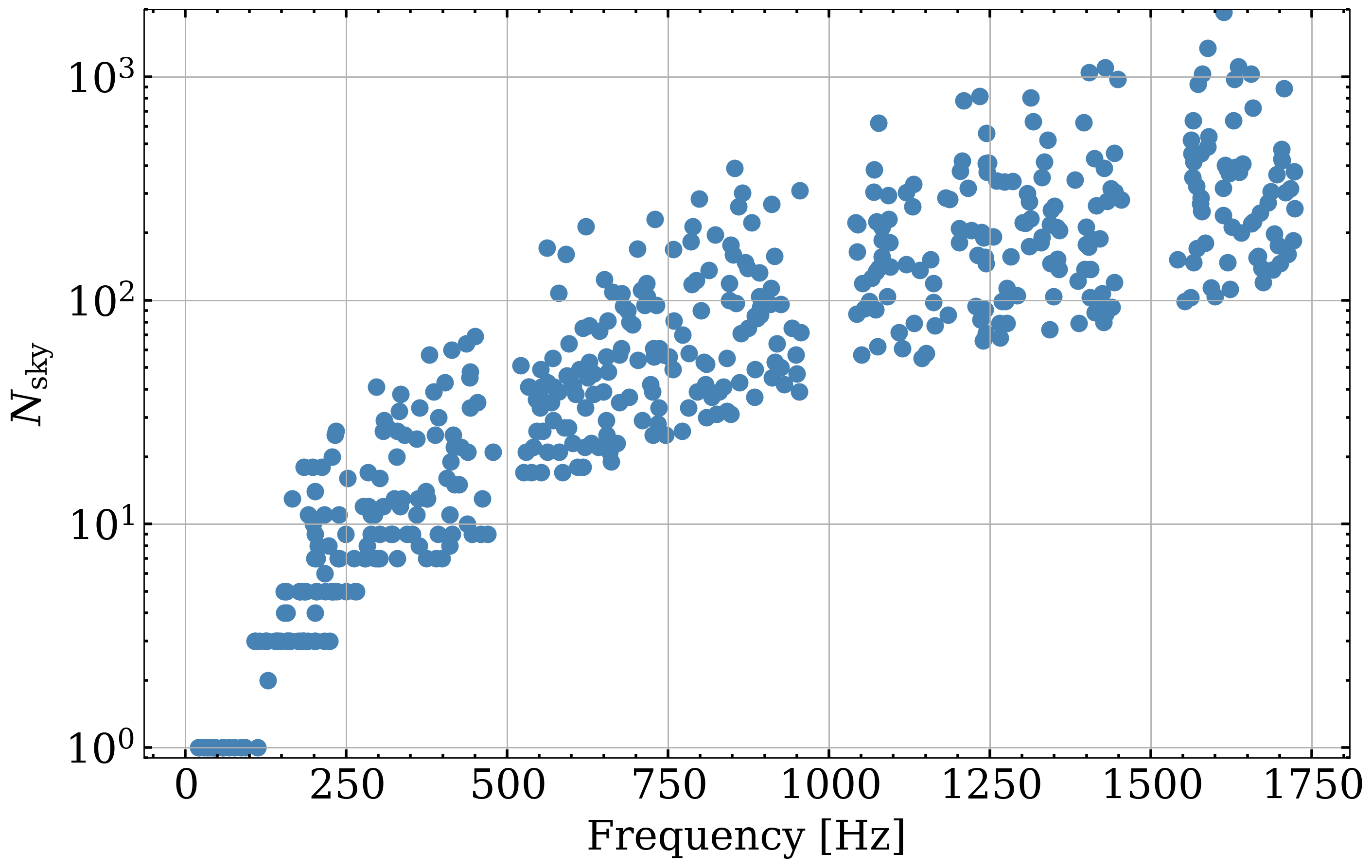}
\caption{\label{fig:templates} Number of searched sky templates, $N_{\rm sky}$, versus central sub-band frequency for all ASAF candidates. The number of templates scales with frequency squared, but also varies with the sky position. The total number of $(\alpha,\delta,\dot{f})$ templates is given by $N_{\rm tot}=N_\mathrm{sky} N_{\dot{f}}$, where $N_{\dot{f}}=11$ for all candidates.}
\end{figure}

Fig.~\ref{fig:skygrid} shows the mismatch contours surrounding an example injection, which lies at the origin of the diamond-shaped sky pixel. Contours corresponding to different mismatch levels are shown by the white curves. We calculate the fiducial sky resolutions $(\Delta\alpha_i^{\rm fid},\Delta\delta_i^{\rm fid})$ for each sample pixel as follows:
\begin{equation}
    \Delta\gamma_i^{\rm fid} = \frac{1}{2}\bigg(\max_{\mu < 0.2}\gamma - \min_{\mu < 0.2}\gamma\bigg)\,,
\end{equation}
where $\gamma\in\{\alpha,\delta\}$, i.e.~we estimate the $\alpha$ and $\delta$ ranges of the sky region enclosed by the $\mu=0.2$ contour, and take half those ranges as the sky resolution. This yields a coarser search grid, as shown by the green crosses in Fig.~\ref{fig:skygrid}, which is built starting from the pixel origin such that a template always lands on the origin.
The $(\Delta\alpha_i^{\rm fid},\Delta\delta_i^{\rm fid})$ values are then interpolated over all $3072$ possible sky pixels. The sky grid for a particular ASAF candidate located at pixel $i$ and central frequency $f_0$ is found through a simple scaling of this fiducial grid,
\begin{equation}
    \Delta\gamma_i = \Delta\gamma_i^{\rm fid}\bigg(\frac{1\,\,\mathrm{kHz}}{f_0}\bigg)\,.
\end{equation}
The computational cost of the search scales with the square of the frequency, owing to the greater number of sky templates needed to guarantee a reasonable mismatch. The number of searched sky templates, $N_{\rm sky}$, for all ASAF candidates is shown in Fig.~\ref{fig:templates}, and ranges from a single sky position for $f_0<120\,\,\mathrm{Hz}$, up to $N_{\rm sky}\sim 10^3$ for the highest-frequency candidates.

\subsubsection{Mismatches} \label{sec:mismatches}

\begin{figure}
\centering
\includegraphics[width=1\columnwidth]{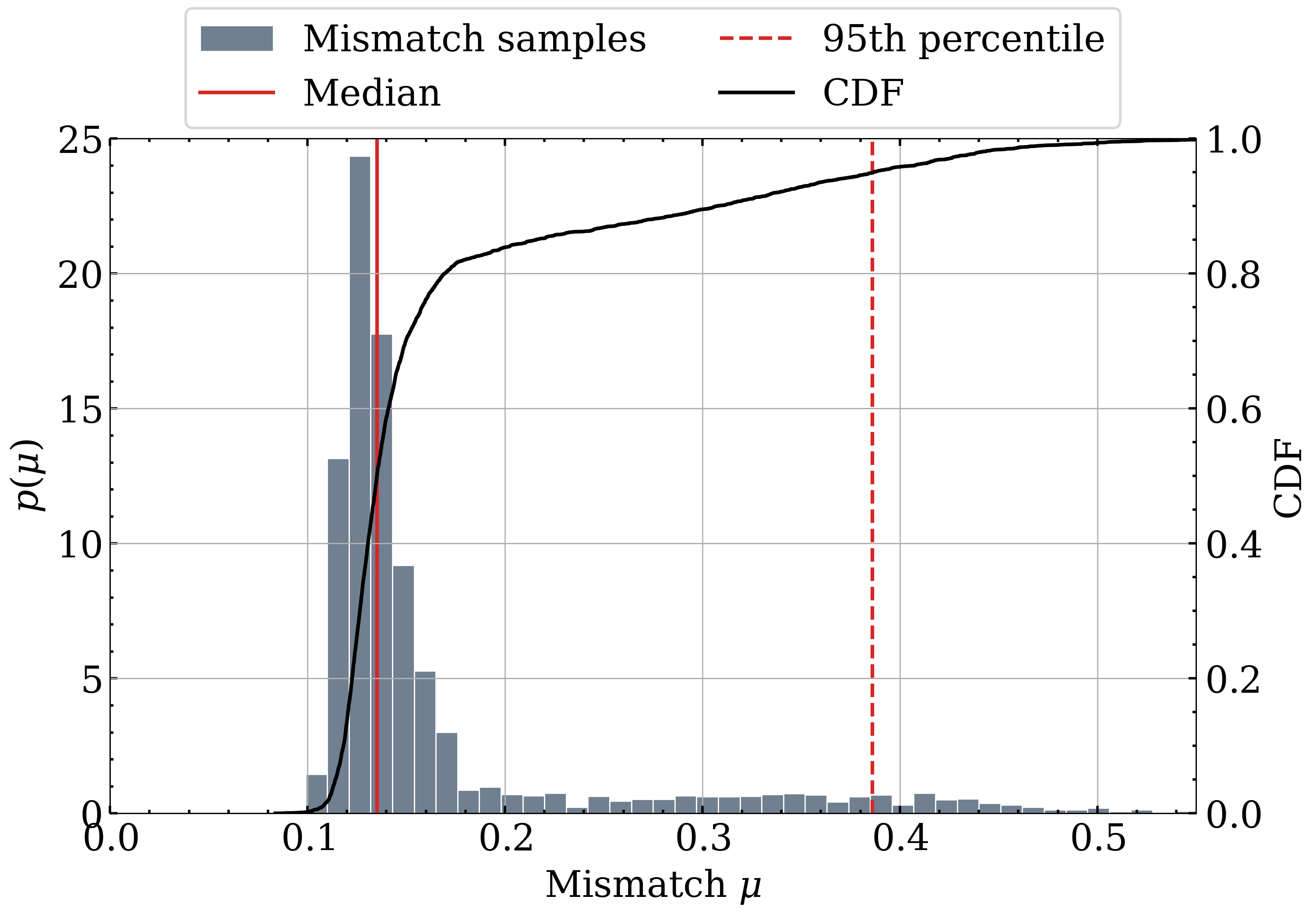}
\caption{\label{fig:mismatch} Probability distribution function (PDF), $p(\mu)$, and cumulative distribution function (CDF) of the mismatch, obtained using $4000$ injection-recovery tests to validate the template grids. The PDF is represented by the grey histogram bins with the scale given on the left axis, while the CDF is the solid black curve with the scale on the right axis. The solid and dashed vertical lines indicate the median and $95$th percentile mismatch values, respectively. The tail of larger mismatches is due to circumstances in which the injected signal simultaneously falls far from both a $\dot{f}$ and sky position template.}
\end{figure}

We verify that our template grids provide an acceptable range of mismatches by simulating the search on $200$ mock ASAF candidates with random sky pixels and $1/32\,\,\mathrm{Hz}$ frequency sub-bands. For each mock candidate, we perform $20$ separate injection-recovery tests, in which a signal is injected into synthetic Gaussian noise in H1 and L1 (fixing $h_0/\sqrt{S_{\rm n}}=1$) with frequency and sky position sampled uniformly within the sub-band and sky pixel, and $\dot{f}$ sampled uniformly from $\pm 10^{-9}\,\,\mathrm{Hz/s}$. We analyze the simulated data twice---first by searching with the empirical template grid, and then searching at a template with identical parameters to the injection, which lets us compute a mismatch via Eq.~\ref{eq:mismatch}. To reduce the computational burden, we only search templates within one parameter grid step from the injection.\footnote{Rarely, when the candidate is at low frequency ($f_0 \lesssim 100\,\,\mathrm{Hz}$) and located near the celestial poles, and a signal is injected near the edge of the sky pixel (inclusive), there are no templates within one grid step of the injection. This occurs because of the peculiar pixel shapes near the poles and down-scaled template density from searching at low frequency. In such cases, we expand the search to include templates within two grid steps from the injection.} The probability distribution function of the mismatch, $p(\mu)$, between the most significant grid template and the injection-matching template is shown in Fig.~\ref{fig:mismatch}. The median is $\mu=0.135$, and is constrained below $\mu=0.386$ at the $95$th percentile. Although the sky grids are constructed via the $\mu=0.2$ mismatch contour, the empirical mismatch distribution extends to $\mu>0.2$ at the tail, which is due to the additional parameter space complexity induced by searching over both $\dot{f}$ and sky position.

\section{Outlier identification} \label{sec:sig}

In this section, we outline our procedure for identifying interesting outliers resulting from our ASAF follow-up. We set empirically derived thresholds for every ASAF candidate by searching detector data at numerous sky positions shifted off-target from the ASAF sky pixel, following the procedures in Refs.~\cite{Middleton:2020skz, LIGOScientific:2021ozr}. Shifting the search off-target allows one to sample the distribution of the detection statistic, $p(\mathcal{L})$, in noise, i.e.~not in the presence of a putative signal. The thresholds, $\mathcal{L}_{\rm th}$, are calculated based on a pre-determined false alarm probability, which we set at $5\%$ per ASAF candidate, i.e.~each candidate has a $5\%$ chance of yielding on outlier due to random noise. For each candidate, we record the loudest log-likelihood template, $\mathcal{L}^*\equiv \max \mathcal{L}$, and compare it to the threshold. If $\mathcal{L}^*>\mathcal{L}_{\rm th}$, then the template is deemed an ``outlier'' and is flagged for further investigation.

\subsection{Thresholds} \label{sec:thresholds}

Historically, HMM CW searches have set thresholds using one of two methods: a parametric approach, where an exponential model is fitted to the tail of the noise distribution \cite{2017PhRvD..96j2006S, LIGOScientific:2019lyx}, and a non-parametric approach, where the threshold is defined to be some percentile of the distribution \citep[e.g.,][]{Middleton:2020skz}. While the latter method is more robust to template correlations and non-Gaussian data, it generally requires a much larger number of off-target simulations to guarantee an accurate percentile. Due to the large computational cost of the number of trials required for the percentile method, we employ the parametric approach, which reduces the computational burden. 

\begin{figure}
\centering
\includegraphics[width=1\columnwidth]{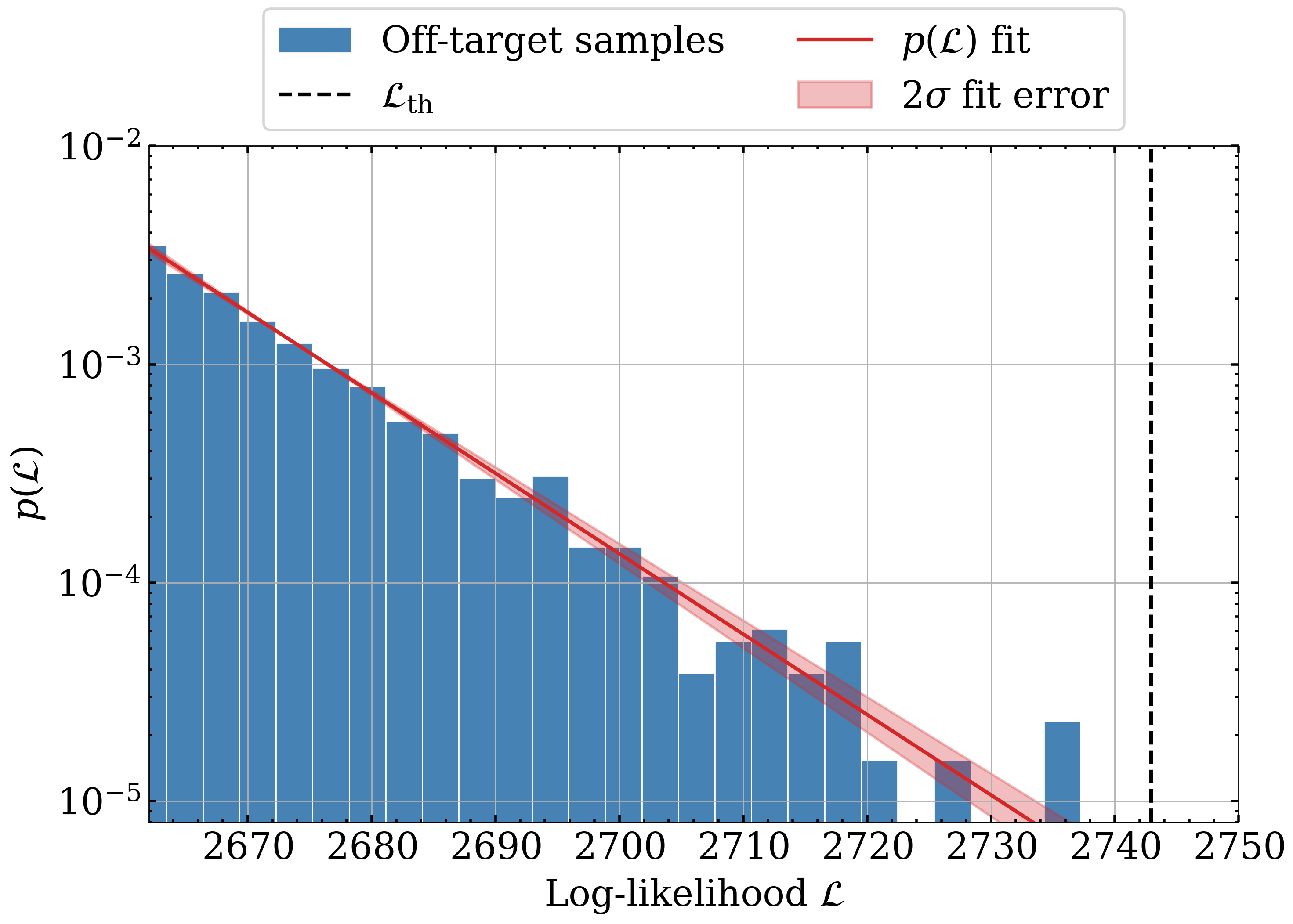}
\caption{\label{fig:threshold} Tail distribution of the log-likelihood statistic, $p(\mathcal{L})$, obtained from off-target simulations around the ASAF candidate at $f_0=708.65625$ Hz, $\alpha_0=4.28$ rad, $\delta_0=-1.11$ rad. The off-target samples are shown by the blue histogram. The solid red line is the maximum likelihood fit using an exponential function. The shaded red region represents the $2\sigma$ fit confidence band. The black dashed line is the threshold calculated using the method in Sec.~\ref{sec:thresholds}.}
\end{figure}

Following Ref.~\cite{LIGOScientific:2021ozr}, we model the tail of the off-target distribution as
\begin{equation} \label{eq:exptail}
    p(\mathcal{L}) = k\lambda e^{-\lambda(\mathcal{L} - \mathcal{L}_{\rm tail})}\,,\quad \mathcal{L} > \mathcal{L}_{\rm tail}\,,
\end{equation}
where $\lambda$ and $\mathcal{L}_{\rm tail}$ are the slope and location parameters, respectively, and $k$ is a normalization factor. Because the detector sensitivity varies as a function of frequency and sky position, we expect the thresholds to vary depending on the location of the ASAF candidate. Thus, we sample and fit (using Eq.~\ref{eq:exptail}) the off-target distributions for all ASAF candidates, deriving for each candidate an independent threshold. In order to obtain samples from the off-target distribution, we search at $N_{\rm off}=4000$ off-target positions from a disjoint sky region around each ASAF candidate,
\begin{equation}
\begin{split}
    \alpha &\in [\alpha_{\min} - 20^\circ, \alpha_{\min} - 5^\circ] \cup [\alpha_{\max} + 5^\circ, \alpha_{\max} + 20^\circ]\,; \\
    \delta &\in [\delta_{\min}, \delta_{\max}]\,,
\end{split}
\end{equation}
where $\alpha_{\min},\alpha_{\max}$ ($\delta_{\min}, \delta_{\max}$) are the minimum and maximum right ascension (declination) values of the template grid. A buffer of $5^\circ$ in right ascension is included to mitigate contamination of the off-target samples by a possible signal in the on-target sky pixel. Every off-target point is searched with the same frequency binning and $\dot{f}$ grid as the on-target search. With $N_{\rm off}=4000$ off-target points, we obtain $M=N_\mathrm{off}N_{\dot{f}}=44000$ sample statistics, $\mathcal{L}_i$, per candidate. The cutoff for the start of the tail, $\mathcal{L}_{\rm tail}$, is defined as the $96$th percentile of the off-target distribution, by which point the tails consistently appear exponential as long as the candidate sub-band does not contain any loud non-Gaussian disturbances.\footnote{Looking ahead to Sec.~\ref{sec:asafresults}, there are at least $11$ candidates where the sub-band contains excessive non-Gaussian noise, causing the off-target distributions to be poorly fit by an exponential model.} This leaves $N_{\rm tail}=0.04M=1760$ samples in the tail to estimate $\lambda$ ($k=0.04$ is already fixed by the tail cutoff), for which we use the maximum likelihood estimator
\begin{equation} \label{eq:lhat}
    \hat{\lambda} = \frac{N_{\rm tail}}{\sum_{i=1}^{N_{\rm tail}} (\mathcal{L}_i-\mathcal{L}_{\rm tail})}\,.
\end{equation}
An example fit to the off-target distribution using Eq.~\ref{eq:lhat} is shown in Fig.~\ref{fig:threshold}. In the Gaussian limit, the expected $2\sigma$ relative error in $\hat\lambda$ is $2N_{\rm tail}^{-1/2}\sim 4.7\%$, which is also shown in Fig.~\ref{fig:threshold} by the shaded region around the best-fit line. This is in agreement with Monte Carlo simulations, which show that $N_{\rm tail}=1760$ guarantees that $\hat\lambda$ is within $5\%$ of the true value at the $95\%$ confidence level. Note that systematic biases could also arise if, for instance, the tail of the off-target distribution is not well-modelled by Eq.~\ref{eq:exptail}, which could happen if the frequency band is polluted by an detector artifact or other non-Gaussianities. We do not try to correct for such biases here, and instead rely on subsequent follow-up (e.g., vetoes) to eliminate outliers arising from non-Gaussian noise.

After fitting the off-target distribution for a given ASAF candidate, we calculate the false alarm probability per template, i.e.~the probability that any single template is above-threshold, as
\begin{equation} \label{eq:alpha}
    \alpha = \int_{\mathcal{L}_{\rm th}}^\infty p(\mathcal{L})\,{\rm d}{\mathcal{L}} = ke^{-\hat{\lambda}(\mathcal{L_{\rm th}} - \mathcal{L}_{\rm tail})}\,.
\end{equation}
Individual templates are required to reach a higher log-likelihood threshold due to a trials factor caused by searching multiple templates per candidate. The false alarm probability over a search involving $N_{\rm tot}=N_\mathrm{sky} N_{\dot{f}}$ statistically independent templates is related to the single-template false alarm probability by
\begin{equation} \label{eq:alphan}
    \alpha_{N_{\rm tot}} = 1-(1-\alpha)^{N_{\rm tot}}\,.
\end{equation}
We adopt $\alpha_{N_{\rm tot}}=0.05$ as the target false alarm threshold per candidate in our analysis, meaning we should expect to accumulate $0.05\times 515 \sim 26$ false alarms resulting from random noise across all ASAF candidates. In practice, we invert Eq.~\ref{eq:alphan} to solve for $\alpha$, and then compute\footnote{Since the threshold depends on a fit parameter ($\hat\lambda$), one could ask whether the threshold itself has some uncertainty. Ultimately, the log-likelihood threshold that one uses to discriminate between ``interesting'' and ``uninteresting'' outliers is a cutoff with no uncertainty. What is uncertain is the false alarm probability that one believes their threshold corresponds to. That is, the chosen threshold corresponds to a false alarm probability close to, but not exactly the same as the target false alarm probability, $\alpha_{N_{\rm tot}}$.} $\mathcal{L}_{\rm th}$ via Eq.~\ref{eq:alpha},
\begin{equation} \label{eq:thres}
    \mathcal{L}_{\rm th} = -\frac{1}{\hat{\lambda}}\log(\alpha/k) + \mathcal{L}_{\rm tail}\,.
\end{equation}
Given an outlier with log-likelihood $\mathcal{L}^*$ from a search with $N_{\rm tot}$ templates, we can also calculate the probability of obtaining a higher log-likelihood $\mathcal{L}>\mathcal{L}^*$ due to random noise as
\begin{equation} \label{eq:pnoise}
    p_{\rm noise} = 1 - [1 - ke^{-\hat{\lambda}(\mathcal{L}^* - \mathcal{L}_{\rm tail})}]^{N_{\rm tot}}\,.
\end{equation}
The quantity $p_{\rm noise}$ is convenient for comparing results from different ASAF candidates, is it takes into account the trials factor, $N_{\rm tot}$, which differs between candidates. Note that Eq.~\ref{eq:alphan} assumes the templates are uncorrelated. However, some template correlations are expected, as the template grid has been deliberately tuned to prevent large mismatches (Sec.~\ref{sec:grid}). The effective number of independent templates is difficult to compute, owing to the large number of candidates to follow-up and wide parameter ranges, and is not done here. Future studies are needed to develop ways of mitigating or correcting for correlated templates in the context of HMM CW searches, e.g.~by following the method of Ref.~\cite{Tenorio:2021wad}.

Because our targets in this search are informed by the results of the ASAF analysis, our search is not a blind search, which incurs an additional trials penalty. The problem of rigorously quantifying how each stage of a hierarchical analysis affects CW detection confidence is challenging to address either empirically (due to computational limitations) or analytically (due to each analysis employing different detection statistics). This is ultimately outside the scope of this work, so we report trials factors separately from the ASAF analysis. In Appendix \ref{app:asaf}, we examine whether the ASAF detection statistic is correlated with $\mathcal{L}^*$ or $p_{\rm noise}$. In summary, we find no correlation with $\mathcal{L}^*$ and a possible correlation with $p_{\rm noise}$.

\subsection{Veto procedure} \label{sec:vetoes}

We apply three separate vetoes to any outliers resulting from our search: the known lines veto, single interferometer veto, and DM-off veto. These vetoes compare the observed properties of an outlier to those expected of non-Gaussian detector artifacts, and have commonly been used in HMM CW searches \citep[e.g.,][]{2017PhRvD..95l2003A, LIGOScientific:2019lyx, KAGRA:2022dqk, LIGOScientific:2021ozr, Middleton:2020skz}. Outliers which do not pass any of the vetoes are likely to be of instrumental origin and are discarded. We briefly summarize the three vetoes here.

\subsubsection{Known lines veto}

The sensitive band of the Advanced LIGO detectors contains numerous documented spectral artifacts of instrumental origin \cite{davis2021ligo}. These artifacts can introduce excess power in the data that is not always well-modelled by background/off-target estimates, resulting in spurious signals. The known lines veto \cite{2017PhRvD..95l2003A} checks whether the frequency path of the outlier, $f^*(t)$, overlaps with any of the vetted\footnote{``Vetted'' means the line is most likely non-astrophysical, whether or not an exact cause has been identified.} lines compiled in Ref.~\cite{knownlines}, after accounting for Doppler line broadening due to the Earth's motion. Specifically, the outlier is vetoed if its frequency path satisfies
\begin{equation} \label{eq:linesveto}
    |f^*(t_n)-f_{\rm line}| < \frac{v_\oplus}{c}f_{\rm line}\,,
\end{equation}
for any time $t_n$, where $f_{\rm line}$ is the line frequency\footnote{The vetted lines also have intrinsic widths, which are sometimes asymmetric about $f_{\rm line}$. These widths are simply added to the left or right bounds of Eq.~\ref{eq:linesveto}.}, and $v_\oplus$ is the Earth's average orbital speed.

\subsubsection{Single interferometer veto}

Outliers representing true astrophysical signals should be present at both detectors. Thus, repeating the search on data from just one detector should naturally decrease its detection statistic. On the other hand, if the outlier arises from an instrumental artifact affecting one of the detectors, then a single-detector analysis should increase the outlier's detection statistic in the detector where it is present, while decreasing it in the detector where it is absent. This motivates the single interferometer veto, where the outlier template is searched at each detector individually, giving two new log-likelihoods $\mathcal{L}_a<\mathcal{L}_b$. As in Ref.~\cite{2017PhRvD..95l2003A}, the outlier is vetoed if the following conditions are all true: (a) the smaller single-detector statistic is sub-threshold, $\mathcal{L}_a<\mathcal{L}_{\rm th}$, (b) the other is above the (two-detector) log-likelihood, $\mathcal{L}_b>\mathcal{L}^*$, and (c) the frequency path associated with $\mathcal{L}_b$ overlaps with the outlier frequency path, i.e.
\begin{equation} \label{eq:ifoveto}
    |f_b(t_n)-f^*(t_n)| < \frac{v_\oplus}{c}f^*(t_n)\,,
\end{equation}
for any $t_n$. If the outlier satisfies Eq.~\ref{eq:ifoveto}, then it is likely associated with noise at one of the detectors and is vetoed. Otherwise, the behaviour is consistent with either a weak signal, or a noise artifact common to both detectors, and cannot be vetoed. 

\subsubsection{DM-off veto}

Instrumental artifacts are not expected to follow the Doppler modulation patterns of real signals originating from a particular point on the sky. The DM-off veto checks how disabling sky demodulation within the \scF-statistic computation affects the log-likelihood of the outlier \cite{Zhu:2017ujz, 2022PhRvD.106l3011J}. If the log-likelihood increases, $\mathcal{L}_{\rm DM-off} > \mathcal{L}^*$, and the recovered frequency path is still overlapping with the original, DM-on frequency path (as determined by Eq.~\ref{eq:ifoveto}), then the outlier is likely noise and is vetoed.

\section{Search results} \label{sec:results}

We present the results of our follow-up of all 515 ASAF candidates in Sec.~\ref{sec:asafresults}. Our searches yield $14$ outliers that survive all vetoes. In Sec.~\ref{sec:outliers}, we perform deeper follow-up of these outliers by searching with a $T_{\rm coh}=2\,\,\mathrm{d}$ coherence time, ultimately finding no strong evidence that these outliers represent astrophysical signals. We remind the reader that, as discussed in Sec.~\ref{sec:asafresults}, our chosen probability of false alarm per ASAF candidate is $5\%$. Thus, $14$ outliers from $515$ candidates is statistically congruent with all of them being false alarms. We also present sensitivity estimates with respect to isolated and long-period binary sources in Sec.~\ref{sec:sens}.

\subsection{ASAF candidates} \label{sec:asafresults}

\begin{figure*}
\includegraphics[width=2.05\columnwidth]{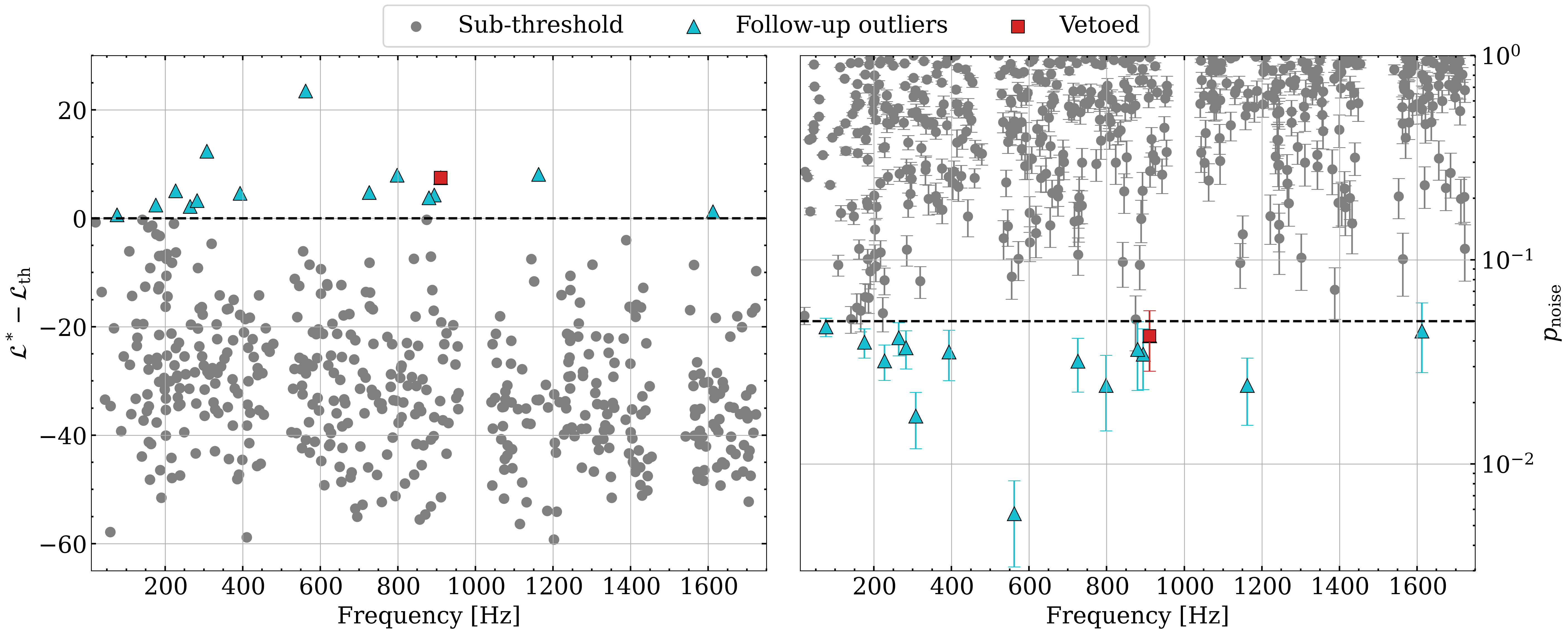}
\caption{\label{fig:results} Follow-up search results for the ASAF candidates. The scatter points correspond to the maximum log-likelihood template obtained for each ASAF candidate. In the left panel, we show the log-likelihood difference $\mathcal{L}^*-\mathcal{L}_{\rm th}$ on the vertical axis, where $\mathcal{L}^*$ is the template log-likelihood and $\mathcal{L}_{\rm th}$ is the threshold. The central frequency of the ASAF candidate is given on the horizontal axis. Templates which surpass the threshold, indicated by the cyan triangles, are identified as outliers and flagged for further follow-up. Sub-threshold templates are in grey. There are a total of 15 outliers, of which one (red square at ${\approx}910$ Hz) is vetoed. In the right panel, the vertical axis is replaced with $p_{\rm noise}$ for each template, with error bars indicating the $2\sigma$ statistical uncertainty propagated from the fit to the off-target distribution, i.e.~they do not include systematics related to template correlations or data calibration. The dashed horizontal line is the $5\%$ false alarm threshold. Note that the left panel does not include error bars, as log-likelihood statistics (and the thresholds) are exact quantities.}
\end{figure*}

Fig.~\ref{fig:results} summarizes our search results; for each ASAF candidate, we show the maximum log-likelihood template, highlighting $15$ outliers which pass the log-likelihood threshold corresponding to a $5\%$ false alarm probability per candidate. Table \ref{tab:outliers} lists the outliers and their properties. All but one of these outliers pass the three vetoes listed in Sec.~\ref{sec:vetoes}. The single vetoed outlier terminating at $910.0801\,\,\mathrm{Hz}$ overlaps with the third harmonic of a H1 beam splitter violin mode. 

Both Fig.~\ref{fig:results} and Table \ref{tab:outliers} include the values of $p_{\rm noise}$ calculated with Eq.~\ref{eq:pnoise}. We obtain the smallest $p_{\rm noise}$ value for the outlier terminating at $561.67839\,\,\mathrm{Hz}$, for which we estimate the false alarm probability at $p_{\rm noise}=0.006$. Given that we search $515$ candidates, we already expect to see ${\sim}3$ outliers at this level of significance. If we set a global threshold at $5\%$ across all candidates, then the per-candidate false alarm probability would be $1-(1-0.05)^{1/515} = 9.96\times 10^{-5}$. As we have no outliers with $p_{\rm noise}$ less than this threshold, we determine that none of the outliers are statistically significant. Nonetheless, the $14$ surviving outliers are subjected to further signal consistency tests in Sec.~\ref{sec:outliers}. The optimal frequency evolution recovered by the Viterbi algorithm is shown for each outlier in Appendix \ref{app:outlierfull}.

Note that $11$ of the sub-threshold points are off-scale in Fig.~\ref{fig:results}, all of which correspond to low-frequency sub-bands polluted by non-Gaussian instrument noise. In these cases, the tail of the off-target distribution is poorly fit by an exponential function, and the resulting threshold is unreliable.

\begin{table*}
\caption{\label{tab:outliers} Properties of the $15$ outliers recovered by our search. The first three columns pertain to the ASAF candidate: the central frequency of the $1/32\,\,\mathrm{Hz}$ sub-band, $f_0$, and the central right ascension and declination coordinates of the sky pixel, $(\alpha_0,\delta_0)$. The fourth column gives the total number of $(\alpha,\delta,\dot{f})$ templates searched for that candidate, $N_{\rm tot}$, and the fifth column is the log-likelihood threshold, $\mathcal{L}_{\rm th}$. The next six columns give the properties of the maximum log-likelihood search template: sky position, $(\alpha,\delta)$; frequency derivative, $\dot{f}$; terminating frequency of the demodulated Viterbi path, $f(t_{N_T})$, Viterbi log-likelihood, $\mathcal{L}^*$, and the probability that a search of the ASAF candidate produces an outlier with $\mathcal{L}^*>\mathcal{L}_{\rm th}$ from noise, $p_{\rm noise}$. The last column shows whether (Y) or not (N) the outlier is vetoed. All numerical $(\alpha,\delta,\dot{f})$ values are rounded to the first significant digit of the grid spacing.}\vspace{0.2cm}
\begin{ruledtabular}
\begin{tabular}{l l l l l p{0.5cm} l l c l l l c} 
$f_0$ [Hz] & $\alpha_0$ [rad] & $\delta_0$ [rad] & $N_{\rm tot}$ & $\mathcal{L}_{\rm th}$ & & $\alpha$ [rad] & $\delta$ [rad] & $\dot{f}$ [$10^{-10}$ Hz/s] & $f(t_{N_T})$ [Hz] & $\mathcal{L}^*$ & $p_{\rm noise}$ & Vetoed \\ [0.5ex] 
\hline \noalign{\vskip 0.1cm}
$76.1875$ & $0.88$ & $0.48$ & $11$ & $2634.7$ & & $0.88$ & $0.48$ & $4$ & $76.20042$ & $2635.4$ & $0.047$ & N \\
$175.59375$ & $0.44$ & $0.17$ & $33$ & $2650.3$ & & $0.41$ & $0.17$ & $-6$ & $175.58405$ & $2652.8$ & $0.04$ & N \\
$227.0625$ & $2.65$ & $-0.30$ & $55$ & $2652.2$ & & $2.60$ & $-0.30$ & $-8$ & $227.07017$ & $2657.3$ & $0.032$ & N \\
$264.0$ & $0.83$ & $0.08$ & $55$ & $2657.9$ & & $0.80$ & $0.08$ & $-2$ & $263.98836$ & $2660.1$ & $0.042$ & N \\
$282.3125$ & $4.81$ & $-0.68$ & $88$ & $2655.7$ & & $4.79$ & $-0.70$ & $-6$ & $282.30154$ & $2659.0$ & $0.037$ & N \\
$307.65625$ & $3.93$ & $-1.21$ & $286$ & $2738.5$ & & $3.89$ & $-1.24$ & $0$ & $307.66895$ & $2750.9$ & $0.017$ & N \\
$393.28125$ & $0.93$ & $1.00$ & $330$ & $2717.1$ & & $0.99$ & $1.00$ & $6$ & $393.27929$ & $2721.8$ & $0.035$ & N \\
$561.6875$ & $5.30$ & $1.366$ & $1892$ & $2782.7$ & & $5.34$ & $1.383$ & $-2$ & $561.67839$ & $2806.2$ & $0.006$ & N \\
$725.8125$ & $4.173$ & $-0.43$ & $429$ & $2659.5$ & & $4.180$ & $-0.46$ & $-6$ & $725.8251$ & $2664.2$ & $0.032$ & N \\
$798.25$ & $1.10$ & $-1.315$ & $3124$ & $2782.0$ & & $1.22$ & $-1.339$ & $-10$ & $798.26468$ & $2789.9$ & $0.024$ & N \\
$879.75$ & $3.75$ & $1.107$ & $2453$ & $2741.1$ & & $3.77$ & $1.091$ & $-10$ & $879.75653$ & $2744.8$ & $0.036$ & N \\
$893.78125$ & $0.246$ & $-0.623$ & $1023$ & $2673.9$ & & $0.246$ & $-0.582$ & $8$ & $893.78092$ & $2678.2$ & $0.035$ & N \\
$910.09375$ & $4.566$ & $0.524$ & $1243$ & $3470.2$ & & $4.546$ & $0.537$ & $10$ & $910.0801$ & $3477.7$ & $0.042$ & Y \\
$1162.34375$ & $3.780$ & $-0.524$ & $1309$ & $2672.6$ & & $3.763$ & $-0.533$ & $-2$ & $1162.3295$ & $2680.7$ & $0.024$ & N \\
$1612.21875$ & $2.649$ & $-0.572$ & $3498$ & $2686.8$ & & $2.646$ & $-0.563$ & $-2$ & $1612.21734$ & $2688.1$ & $0.045$ & N \\
\end{tabular}
\end{ruledtabular}
\end{table*}

There are six sub-threshold templates where the $2\sigma$ statistical uncertainty in $p_{\rm noise}$ includes the per-candidate false alarm threshold, even though they are not considered outliers by our $\mathcal{L}^*>\mathcal{L}_{\rm th}$ cut-off. The uncertainty in $p_{\rm noise}$ is propagated from the uncertainty in $\hat\lambda$, and reflects the fact that we cannot know exactly what threshold value corresponds to the target false alarm probability $\alpha_{N_{\rm tot}}$ based on a fit to the off-target distribution. While it is arguable that these six barely sub-threshold templates should be treated as outliers, we do not consider it worthwhile to do so, owing to their doubly sub-threshold status (a sub-threshold outlier associated with a sub-threshold ASAF candidate).

\subsection{Outlier analysis} \label{sec:outliers}

We further investigate the $14$ post-veto surviving outliers by re-analyzing each ASAF candidate that yielded an outlier with twice the original coherence time, $T_{\rm coh}=2\,\,\mathrm{d}$. In accordance with the $T_{\rm coh}$ scaling laws \cite{Riles:2022wwz}, we also double the number of frequency bins and increase the number of $(\alpha,\delta,\dot{f})$ templates by a factor of $16$.\footnote{The $\dot{f}$ spacing goes as $\Delta\dot{f}=\Delta f/T_{\rm coh}$, so doubling the coherence time and number of frequency bins leads to a factor of four increase in the number of $\dot{f}$ templates. Similarly, the spacing in each sky coordinate scales inversely with $T_{\rm coh}$, leading to a factor of four increase in the number of sky grid points. Thus, doubling the coherence time results in a factor of $16$ more $(\alpha,\delta,\dot{f})$ templates.} It is worth noting that increasing $T_{\rm coh}$ unavoidably changes the signal model by limiting the allowed amount of spin-wandering. However, given that $T_{\rm coh}=1\,\,\mathrm{d}$ is already much shorter than the expected spin-wandering timescale of known neutron stars (see Sec.~\ref{sec:hmm}), there is some allowance for doubling $T_{\rm coh}$ from an astrophysical standpoint. For each outlier, we check whether the maximum log-likelihood template from the $T_{\rm coh}=2\,\,\mathrm{d}$ search is no more than one grid step away (in sky position or $\dot{f}$) from the loudest $T_{\rm coh}=1\,\,\mathrm{d}$ template, where the grid step size corresponds to the $T_{\rm coh}=1\,\,\mathrm{d}$ template spacing. None of the outliers satisfy this closeness criterion; for $12$ of them, the sky positions differ by more than one grid step, and, for the other two, the template $\dot{f}$ values differ. 

Simply comparing the loudest templates does not conclusively rule out these outliers as noise, especially at marginal significance, as the $T_{\rm coh}=2\,\,\mathrm{d}$ search could reveal a noise fluctuation which shifts the maximum likelihood to a different template. Thus, it is useful to examine the Viterbi scores, defined as \cite{LIGOScientific:2019lyx, Middleton:2020skz}
\begin{equation}
    \mathcal{S} = \frac{\mathcal{L}_{q^*}-\mu_\mathcal{L}}{\sigma_\mathcal{L}}\,,
\end{equation}
where $\mu_\mathcal{L}$ and $\sigma^2_\mathcal{L}$ denote the log-likelihood mean and variance over all $N_Q$ terminating paths for a single $(\alpha,\delta,\dot{f})$ template,
\begin{equation}
\begin{split}
    \mu_\mathcal{L} &= \frac{1}{N_Q}\sum_{i=1}^{N_Q} \mathcal{L}_{q_i}\,, \\
    \sigma^2_\mathcal{L} &= \frac{1}{N_Q}\sum_{i=1}^{N_Q} (\mathcal{L}_{q_i} - \mu_\mathcal{L})^2\,.
\end{split}
\end{equation}
Here, $\mathcal{L}_{q_i}$ is the log-likelihood of the path ending in frequency bin $q_i$, and $\mathcal{L}_{q^*}$ is the maximum log-likelihood path (Eq.~\ref{eq:loglike}). Unlike the log-likelihood, which depends on the number of steps $N_T$ in the HMM, the score measures the significance of a given path relative to all other frequency paths in the same band, giving a normalized quantity that can be directly compared between searches with different $T_{\rm coh}$. For $T_{\rm coh}=1\,\,\mathrm{d}$, we record the score corresponding to the maximum log-likelihood template, $\mathcal{S}_{\rm 1d}$, as before. For $T_{\rm coh}=2\,\,\mathrm{d}$, we select the score of the maximum log-likelihood template, considering only templates which lie within on grid step from the loudest $T_{\rm coh}=1\,\,\mathrm{d}$ template, where again the grid step size corresponds to the $T_{\rm coh}=1\,\,\mathrm{d}$ search. If the original $T_{\rm coh}=1\,\,\mathrm{d}$ template represents a real signal that is loud enough to be detected, then we expect a more sensitive search over the nearby parameter space to recover the same signal at higher significance, i.e.~with larger Viterbi score. 

As a baseline comparison, we perform $T_{\rm coh}=1\,\,\mathrm{d}$ and $T_{\rm coh}=2\,\,\mathrm{d}$ analyses on synthetic data consisting of either Gaussian noise with injected signals or pure Gaussian noise. For these synthetic data searches, we select a $1/32\,\,\mathrm{Hz}$ band centered at $f_0=800\,\,\mathrm{Hz}$ and a sky pixel with origin $(\alpha_0,\delta_0)=(4.271\,\,\mathrm{rad}, -0.34\,\,\mathrm{rad})$, and search it exactly as done for the ASAF candidates. We generate $200$ data sets for each scenario (Gaussian noise, Gaussian noise with an injection) with different realizations of Gaussian noise and injection parameters. The injected signals are sampled similarly to the mismatch study in Sec.~\ref{sec:mismatches}, i.e.~the injections are randomly injected into the sky pixel and sub-band with $\dot{f}$ sampled uniformly between $\pm 10^{-9}\,\,\mathrm{Hz/s}$. The amplitudes, $h_0$, are sampled uniformly around the $T_{\rm coh}=1\,\,\mathrm{d}$ detection limit, $\sqrt{S_{\rm n}}/h_0\in [35, 60]$, as informed by our sensitivity calculations in Sec.~\ref{sec:sens}. 

\begin{figure}
\centering
\includegraphics[width=1\columnwidth]{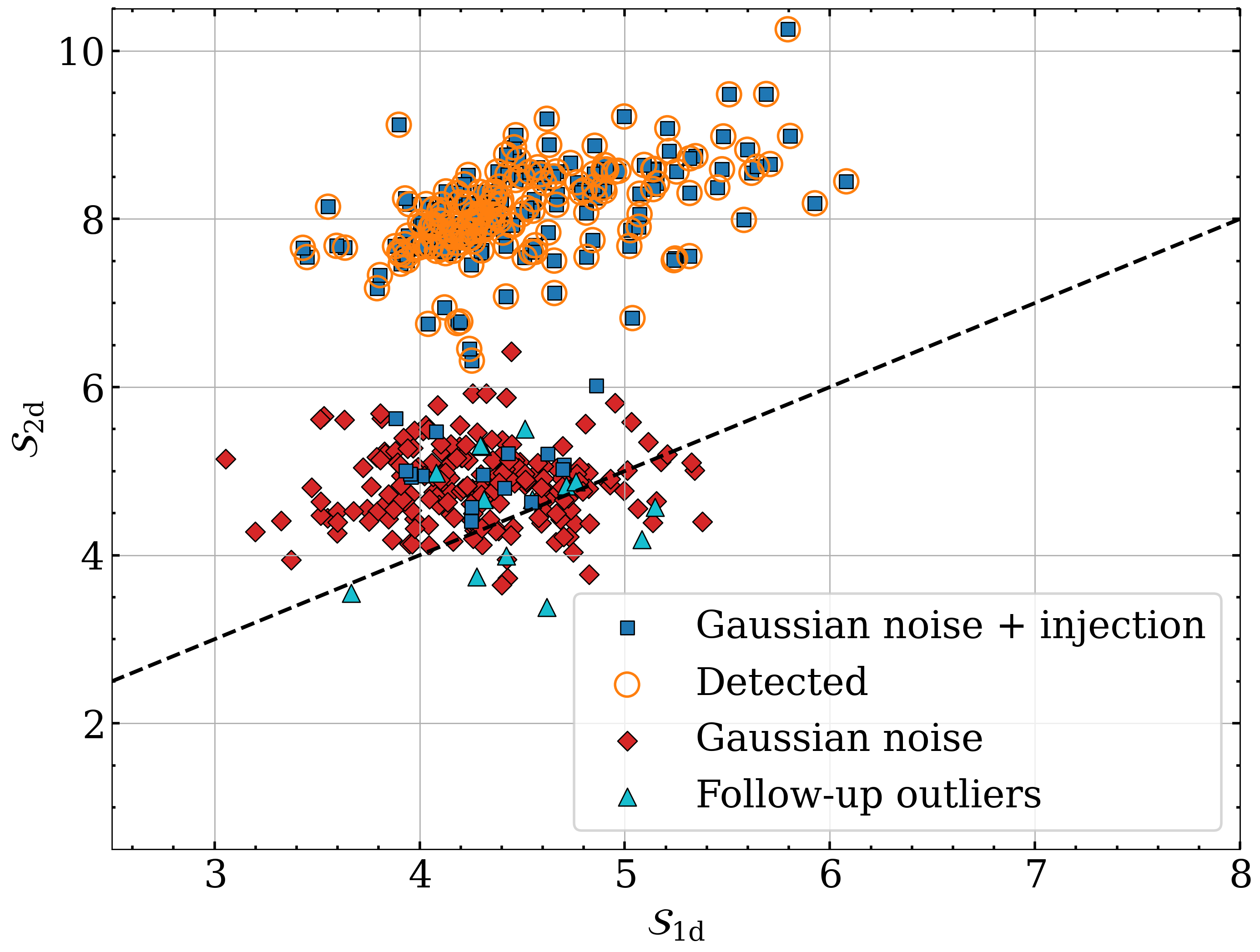}
\caption{\label{fig:followup} Comparison of the Viterbi scores between the $T_{\rm coh}=1\,\,\mathrm{d}$ ($\mathcal{S}_{\rm 1d}$, horizontal axis) and $T_{\rm coh}=2\,\,\mathrm{d}$ ($\mathcal{S}_{\rm 2d}$, vertical axis) searches. The dashed line marks where $\mathcal{S}_{\rm 1d}=\mathcal{S}_{\rm 2d}$. We show results for the 14 surviving outliers (cyan triangles), and for synthetic data containing either pure Gaussian noise (red rhombuses) or Gaussian noise with injected signals (blue squares). Orange circles mark injections which are ``detected'' by both searches, i.e.~the maximum log-likelihood template is within one parameter grid step (assuming the $T_{\rm coh}=1\,\,\mathrm{d}$ template spacing) from the injection parameters. The outliers are broadly consistent with the Gaussian noise scenario. When an injection is too weak to be detected, the scores fall into the Gaussian noise region, as shown by the blue squares which are not circled.}
\end{figure}

We show the results for our injection studies alongside the ASAF follow-up outliers in Fig.~\ref{fig:followup}, where we observe how the scores tend to cluster around different $\mathcal{S}_{\rm 2d}/\mathcal{S}_{\rm 1d}$ ratios depending on whether a signal is detected or not. When the data contain an injected signal, the scores cluster around larger ratios compared to the pure Gaussian noise scenario, as the SNR is amplified relative to the background. We observe that the scores for the follow-up outliers cluster in the same region as the Gaussian noise simulations, suggesting their behaviour is more consistent with Gaussian noise than an astrophysical signal. Taken with their low significance, and the fact that the $T_{\rm coh} = 2\,\,\mathrm{d}$ analyses recover different maximum log-likelihood templates, we conclude that there is no strong evidence in support of the outliers being genuine astrophysical signals.

\subsection{Sensitivity} \label{sec:sens}

We derive empirical sensitivity estimates for our search by recovering simulated signals injected into O3 detector data. The sensitivity is defined as the minimum strain amplitude, $h_0^{95\%}$, at which $95\%$ of all injected signals are recovered above the log-likelihood threshold. Another commonly used metric is the sensitivity depth, which measures the sensitivity relative to the noise level of the detectors \cite{Behnke:2014tma, Dreissigacker:2018afk}:
\begin{equation}
    \mathcal{D}^{95\%} = \frac{\sqrt{S_{\rm n}}}{h_0^{95\%}}\,,
\end{equation}
where $S_{\rm n}$ is the inverse-squared average, running-median PSD for both detectors. We estimate both $h_0^{95\%}$ and $\mathcal{D}^{95\%}$ at $100$ ASAF candidates that contain no outliers or instrument lines. This lets us reuse the previously calculated thresholds for those candidates, without the need for additional off-target simulations.

\subsubsection{Injection campaign}

We characterize our sensitivity against two types of signals: isolated sources, and sources in long-period binary systems with orbital period $P_{\rm b}>1$ yr. For each candidate and source type, we generate $N_{\rm inj}=800$ injections within the candidate sky pixel and sub-band. We further stipulate that the signal frequency must be contained within the $1/32\,\,\mathrm{Hz}$ band. For the binary injections, we thus fix the signal frequency (at the reference time) to the central frequency of the sub-band, and apply a crude Doppler amplitude cut of $2\Delta f_{\rm orb}<1/32\,\,\mathrm{Hz}$, where 
\begin{equation} \label{eq:orbdoppler}
    \Delta f_{\rm orb} = \frac{2\pi a_{\rm p}}{P_{\rm b}} \bigg(\frac{1+e}{1-e}\bigg)^{1/2}\,,
\end{equation}
is the maximum possible Doppler shift caused by the binary orbital motion of the source. This ensures the frequency path of the injected signal does not drift outside the sub-band. In Eq.~\ref{eq:orbdoppler}, $a_{\rm p}$ is the projected semimajor axis along the line-of-sight in units of light-seconds (ls), and $e$ is the orbital eccentricity. The remaining parameters are drawn uniformly in the same manner for both source types: frequency derivative, $\dot{f}\in[-10^{-9}\,\,\mathrm{Hz/s}, 10^{-9}\,\,\mathrm{Hz/s}]$, strain amplitude, $h_0/10^{-25}\in[0.2, 4]$, cosine inclination angle, $\cos\iota\in[-1, 1]$, polarization angle, $\psi\in[-\pi/2,\pi/2]$, and reference phase, $\phi_0\in[0, 2\pi]$. For the binary injections, we also sample log-uniformly over the projected semimajor axis, $a_{\rm p}\in[10\,\,\mathrm{ls}, 10^5\,\,\mathrm{ls}]$, binary orbital period, $P_{\rm b}\in[1\,\,\mathrm{yr}, 10^3\,\,\mathrm{yr}]$, eccentricity, $e\in[10^{-7}, 0.95]$, and uniformly over the time of passage of the ascending node (in GPS time), $t_\mathrm{asc}\in [t_0-10^8\,\,\mathrm{s},t_0+10^8\,\,\mathrm{s}]$, where $t_0=1253764818.0$ is the midpoint GPS time of O3. We run our search algorithm on each injection, searching only the templates which are within one grid step in sky position and $\dot{f}$ from the injection parameters. We do not search over any binary parameters, instead letting the HMM and Viterbi algorithm accommodate the frequency variation induced by the binary motion.

\subsubsection{Sensitivity estimates}

\begin{figure}
\centering
\includegraphics[width=1\columnwidth]{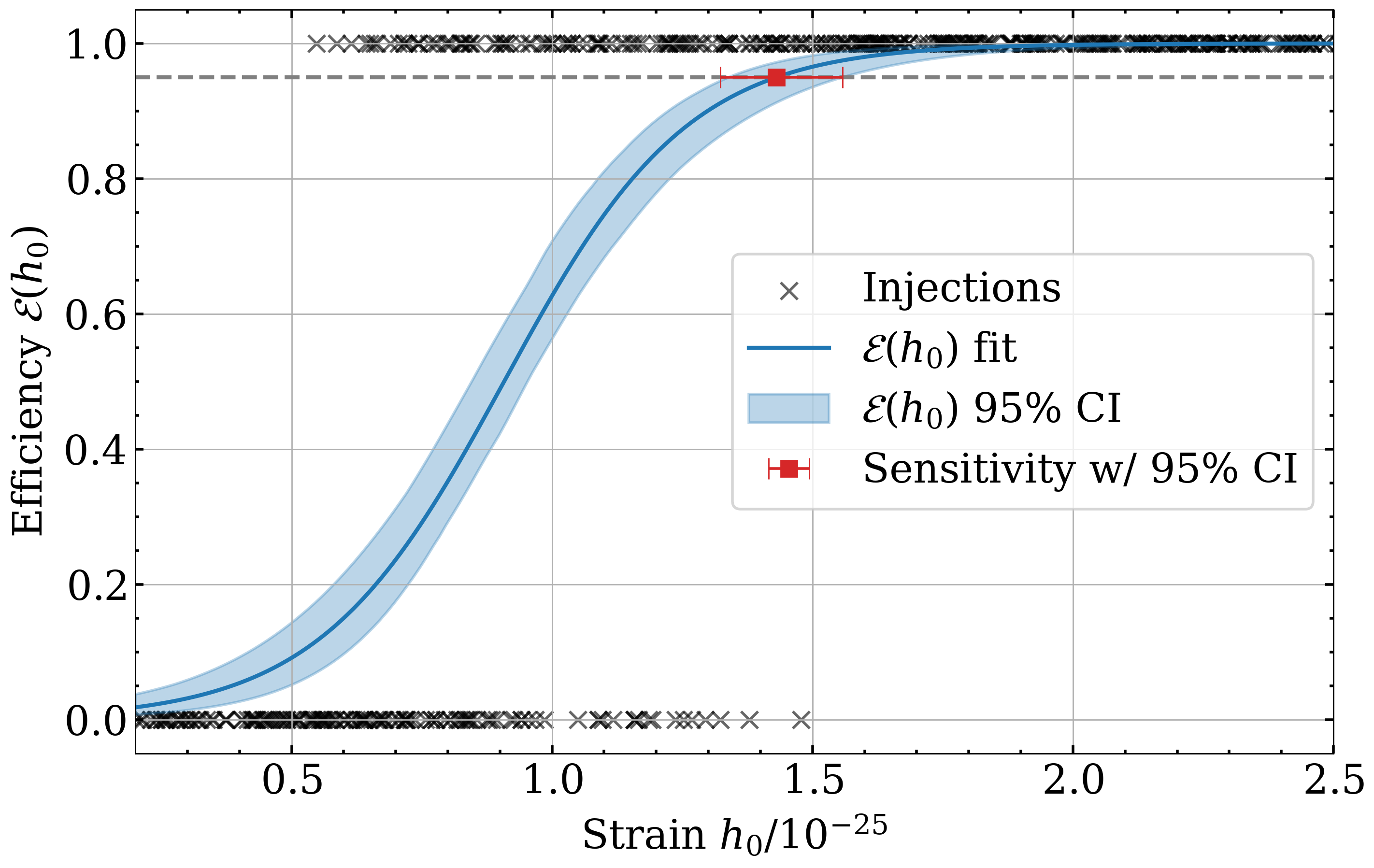}
\caption{\label{fig:logreg} Example sensitivity estimate for an ASAF candidate. Each black cross corresponds to a simulated signal injected into O3 data, which is assigned a value of ``$1$'' if it was detected above threshold, and ``$0$'' otherwise. The binary data are fitted with a logistic regression, letting us estimate where the detection efficiency crosses the $95\%$ threshold (grey line). The solid blue curve is the median posterior fit, with the $95\%$ credible interval (CI) shown by the shaded blue region. The constraint on the sensitivity, $h_0^{95\%}$, is shown by the red square, with error bars corresponding to the $95\%$ CI.}
\end{figure}

The fraction of detected signals at a given strain amplitude, known as the detection efficiency, $\mathcal{E}(h_0)$, is modeled as a logistic function,
\begin{equation} \label{eq:logreg}
    \mathcal{E}(h_0) = \frac{1}{1+e^{-\beta_0-\beta_1 h_0}}\,,
\end{equation}
where $\beta_{0,1}$ are fit parameters. We seek the amplitude $h_0^{95\%}$ such that $\mathcal{E}(h_0^{95\%})=0.95$. First, Eq.~\ref{eq:logreg} must be fitted using the injection search results. This can be done by either binning the injections by $h_0$ and counting the number of detections in each bin, or using binary logistic regression, which avoids the need for binning. We opt for the latter approach, finding the fit parameters which maximize the log-likelihood function:
\begin{equation}
    \ell(\beta_0,\beta_1) = \sum_{i=1}^{N_{\rm inj}}[d_i\log(\mathcal{E}_i) + (1-d_i)\log(1-\mathcal{E}_i)]\,,
\end{equation}
where $h_i$ is the strain amplitude of the $i$th injection, $d_i=1$ if the injection is detected\footnote{An injection is considered ``detected'' if any of the templates in the vicinity of the injection, defined as being within one grid step in $\dot{f}$ and sky position, pass the log-likelihood threshold.} ($d_i=0$ otherwise), and $\mathcal{E}_i=\mathcal{E}(h_i)$. We determine the best fit parameters numerically using Markov chain Monte Carlo (MCMC) sampling, and then transform the joint posterior distribution for $\beta_{0,1}$ to the posterior distribution for $h_0^{95\%}$ using Eq.~\ref{eq:logreg}, assuming flat priors on $\beta_{0,1}$. Fig.~\ref{fig:logreg} shows an example of using this logistic regression technique to estimate $h_0^{95\%}$.

\begin{figure}
\centering
\includegraphics[width=1\columnwidth]{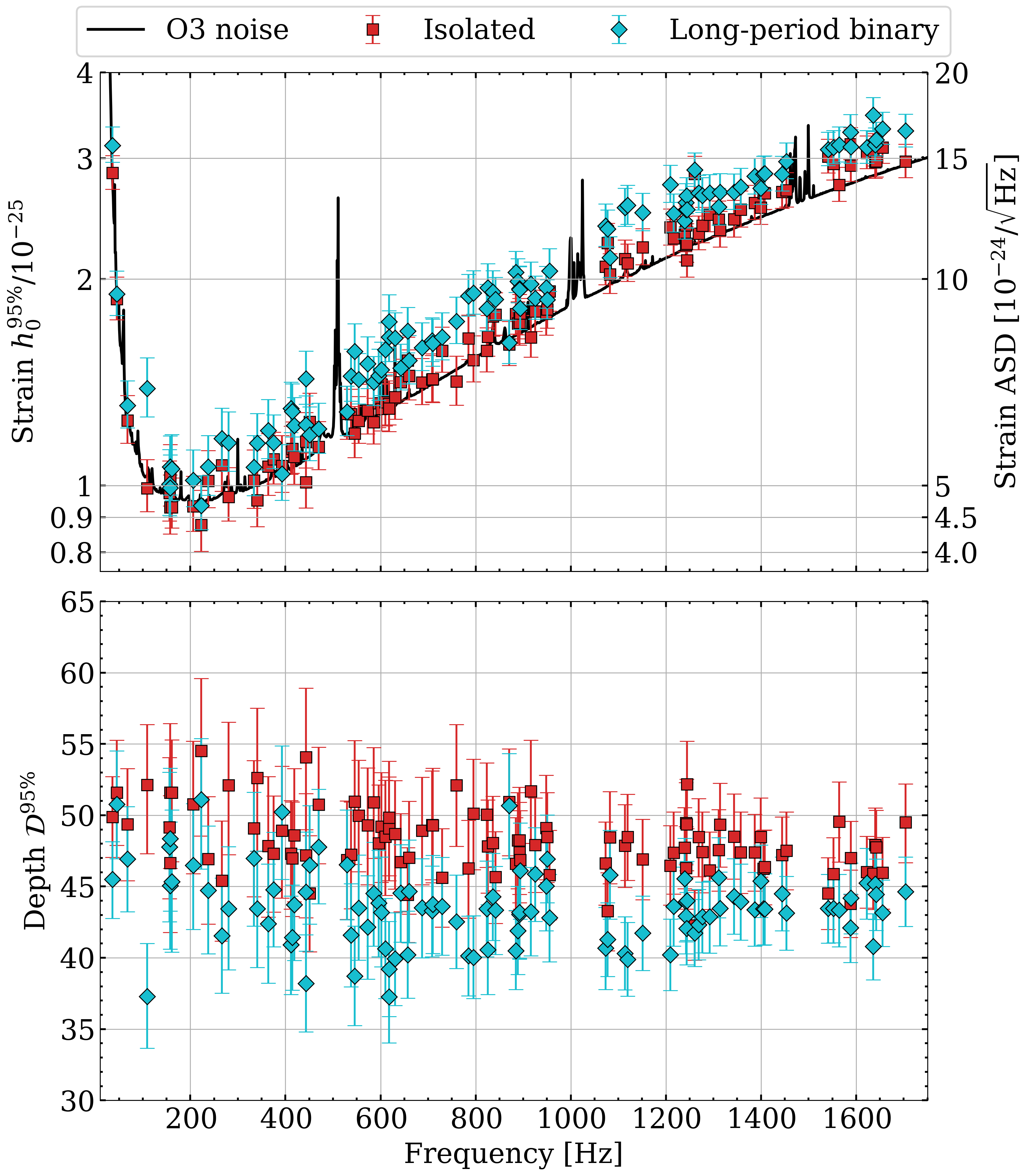}
\caption{\label{fig:sensitivity} Sensitivity limits for our search, estimated for $100$ of the $515$ total ASAF candidates. The colour indicates the sensitivity to isolated sources (red squares) or long-period binaries (cyan rhombuses). In the top panel, we show the minimum strain amplitude, $h_0^{95\%}$, at which we detect $95\%$ of the injections, marginalized over all other source parameters. The horizontal axis gives the central frequency of the candidate. The black curve shows an O3 run-averaged Advanced LIGO noise spectrum, which has been scaled to match the general shape of the sensitivities. In the bottom panel, we show the sensitivity depth for the same set of ASAF candidates. The error bars on the estimated sensitivities correspond to the $95\%$ credible interval from the logistic regression.}
\end{figure}

In Fig.~\ref{fig:sensitivity}, we report sensitivity limits for a representative sample of $100$ ASAF candidates, marginalized over all other source parameters. The highest sensitivity is achieved around $222.6$ Hz, where the minimum detectable strain amplitude is $h_0^{95\%} = 8.8\times 10^{-26}$ for isolated sources and $h_0^{95\%} = 9.4\times 10^{-26}$ for long-period binaries, corresponding to sensitive depths of $55$ and $51$, respectively. We observe that the depth decreases slightly at higher frequencies. Since the number of templates scales with the square of the frequency, the thresholds are higher due to the trials penalty, which reduces the sensitivity of the search. The sensitivity to long-period binaries is consistently below that of isolated sources. This is not unexpected given that we only considered isolated signals when calibrating the mismatch in Sec.~\ref{sec:mismatches}. Overall, our limits are competitive with previous directed HMM CW searches which used comparable coherence times \cite{LIGOScientific:2021mwx}, as well as the minimum upper limits set by O3 all-sky CW searches \cite{KAGRA:2022dwb, Steltner:2023cfk} and the ASAF analysis \cite{KAGRA:2021rmt}. However, we caution that these sensitivities/upper limits are conditional on the assumed signal model and false alarm threshold, and are not directly comparable with each other.

We note that the long-period binary sensitivities reported in this work rely on the signal remaining mostly contained inside the frequency search band. If the amplitude cut (Eq.~\ref{eq:orbdoppler}) or frequency centering assumptions are relaxed, our sensitivity worsens by a factor between $2$--$3$, as the binary motion-induced Doppler modulation can easily shift the signal out of band for CW frequencies $\gtrsim 100\,\,\mathrm{Hz}$. Future studies interested in optimizing Viterbi for use in long-period binary searches should consider using larger frequency bands to accommodate the binary Doppler shift, and/or include long-period binary templates in the matched filter search.

\section{Conclusions} \label{sec:conclusion}

In this study, we performed a triggered CW search by following up sub-threshold stochastic GW candidates. Our CW search uses a HMM framework, combined with the \scF-statistic, to achieve a flexible signal model that is sensitive to a wide range of frequency evolution \cite{Suvorova:2016rdc, 2018PhRvD..97d3013S, KAGRA:2022dqk, LIGOScientific:2021mwx, LIGOScientific:2021ozr, Middleton:2020skz, PhysRevD.102.083025, 2019PhRvD.100b3006B, LIGOScientific:2021rnv, PhysRevD.103.083009, Beniwal:2022nam, Vargas:2022mvs}. Using Advanced LIGO data from O3, we have applied this method to follow-up all $515$ sub-threshold candidates identified in the LVK O1--O3 ASAF analysis \cite{KAGRA:2021rmt}, marking the first time a CW search has targeted stochastic candidates. Our search used a coherence time of $24\,\,\mathrm{hr}$ and $10^{-5}\,\,\mathrm{Hz}$ frequency bins, providing robustness to signals whose frequency changes by at most  $10^{-5}\,\,\mathrm{Hz}$ per day after correcting for secular spin-down/up, which is enough to capture the spin-wandering behaviour of most known neutron stars (e.g., Scorpius X-1 \cite{KAGRA:2022dqk}).


Of the $515$ ASAF candidates, we obtain $15$ outliers of marginal significance passing a threshold corresponding to a $5\%$ false alarm probability per candidate. Applying a set of vetoes eliminates one outlier, leaving $14$ surviving outliers. We then subjected these outliers to additional follow-up by repeating the search with double the coherence time ($48\,\,\mathrm{hr}$). None of our outliers are recovered by this more sensitive, longer-coherence search, nor do they exhibit the increase in log-likelihood expected of CW signals, suggesting the outliers are more consistent with random noise. These investigations lead us to conclude that our outliers are unlikely to represent true astrophysical signals. Data from future observing runs may reveal whether any of the outliers persist, or disappear into the noise background.


The sensitivity of our search to isolated and long-period binary sources is investigated with injection studies. We obtain sensitivity limits comparable with current all-sky CW upper limits and the ASAF upper limits. The minimum detectable strain for isolated sources with no spin-wandering is $8.8\times 10^{-26}$, achieved at a frequency of $222.6$ Hz, and the corresponding sensitivity to long-period binary systems with orbital periods greater than $1$ yr is $9.4\times 10^{-26}$. The HMM approach allows one to search for long-period binaries without having to search over any binary orbital templates, reducing the required number of search templates and associated computational cost.

Our search method has broad applicability beyond following up stochastic candidates, and could be used to follow up high-priority candidates identified by all-sky CW searches. As we assume an alternative, spin-wandering signal model, our method should prove generally flexible to follow up outliers with poor signal phase constraints, and is complementary to currently existing, fully coherent follow-up procedures which assume a constant or secularly evolving signal frequency.

\begin{acknowledgments}
We thank James Clark for assistance with submitting computing jobs through the IGWN Computing Grid. We thank the LVK CW and Anisotropic working groups for helpful discussions. 

This material is based upon work supported by NSF's LIGO Laboratory which is a major facility fully funded by the National Science Foundation. This research was done using services provided by the OSG Consortium \cite{osg07, osg09, osgdoi}, which is supported by the National Science Foundation awards OAC-2030508 and OAC-1836650. This research was enabled in part by support provided by the Digital Research Alliance of Canada (\url{alliance.can.ca}). We are also grateful for computational resources provided by the LIGO Laboratory and supported by National Science Foundation awards PHY-0757058 and PHY-0823459. 
A.M.K., H.D., E.G., and J.M. acknowledge funding support from the Natural Sciences and Engineering Research Council of Canada. 
A.M.K. acknowledges funding support from a Killam Doctoral Scholarship.
J.B.C., L.S., L.D., L.S., and A.M. acknowledge the support of the Australian Research Council Centre of Excellence for Gravitational Wave Discovery (OzGrav), Project No. CE170100004.
H.M. acknowledges the support of the UK Space Agency Grant No.~ST/V002813/1 and ST/X002071/1.

\end{acknowledgments}

\appendix

\section{Outlier search results} \label{app:outlierfull}

More detailed search results for the $15$ outliers, including the Viterbi frequency paths, are reported in Fig.~\ref{fig:outlierfull}. Note that the outlier in Fig.~\ref{fig:outlier_m} is vetoed. Visually inspecting the Viterbi paths, the majority of the outliers exhibit large jumps in frequency before and after the O3 commissioning break (Figs.~\ref{fig:outlier_a}, \ref{fig:outlier_d}, \ref{fig:outlier_e}, \ref{fig:outlier_f}, \ref{fig:outlier_g}, \ref{fig:outlier_h}, \ref{fig:outlier_j}, \ref{fig:outlier_k}, \ref{fig:outlier_l}, \ref{fig:outlier_m}, \ref{fig:outlier_o}). The sizes of these jumps ranges between $25$--$30$ frequency bins, implying the frequency track moved almost exclusively upwards or downwards for many consecutive days. While a frequency jump of this nature is technically allowed by the random walk signal model, the coincidence with the O3 break is certainly suspicious. We ultimately discard all of these outliers based on our investigations in Sec.~\ref{sec:outliers}.

\begin{figure*}
\subfloat[\label{fig:outlier_a} Template: $\alpha=0.88$, $\delta=0.48$, $\dot{f}=4\times 10^{-10}$ Hz/s.]{\includegraphics[width=1.3\columnwidth]{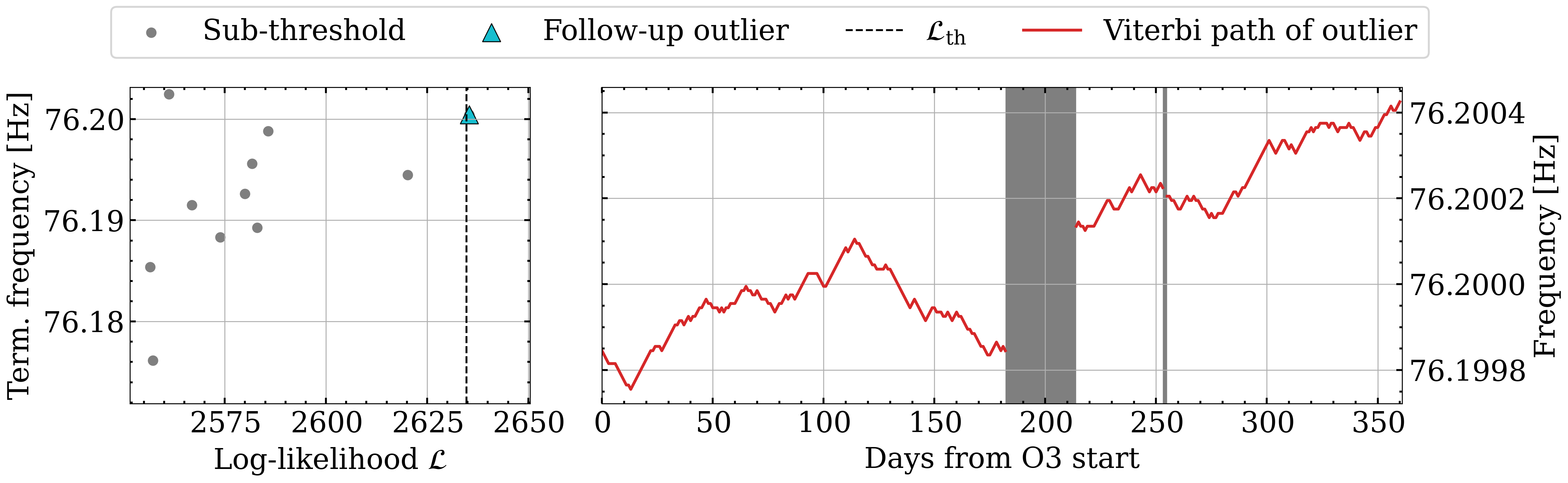}}
\hfill
\subfloat[Template: $\alpha=0.41$, $\delta=0.17$, $\dot{f}=-6\times 10^{-10}$ Hz/s.]{\includegraphics[width=1.3\columnwidth]{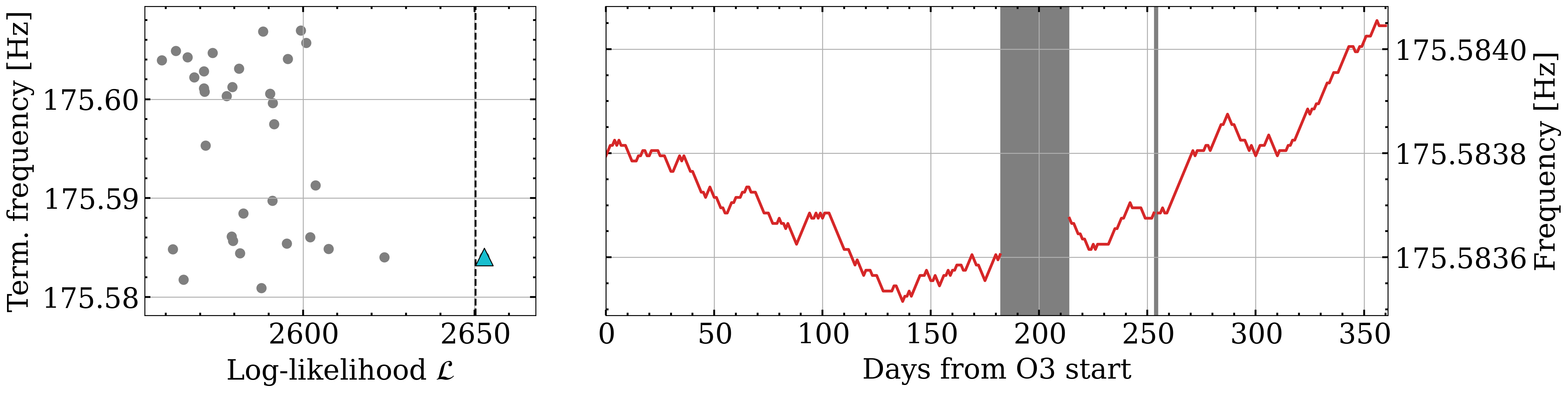}}
\hfill
\subfloat[Template: $\alpha=2.60$, $\delta=-0.30$, $\dot{f}=-8\times 10^{-10}$ Hz/s.]{\includegraphics[width=1.3\columnwidth]{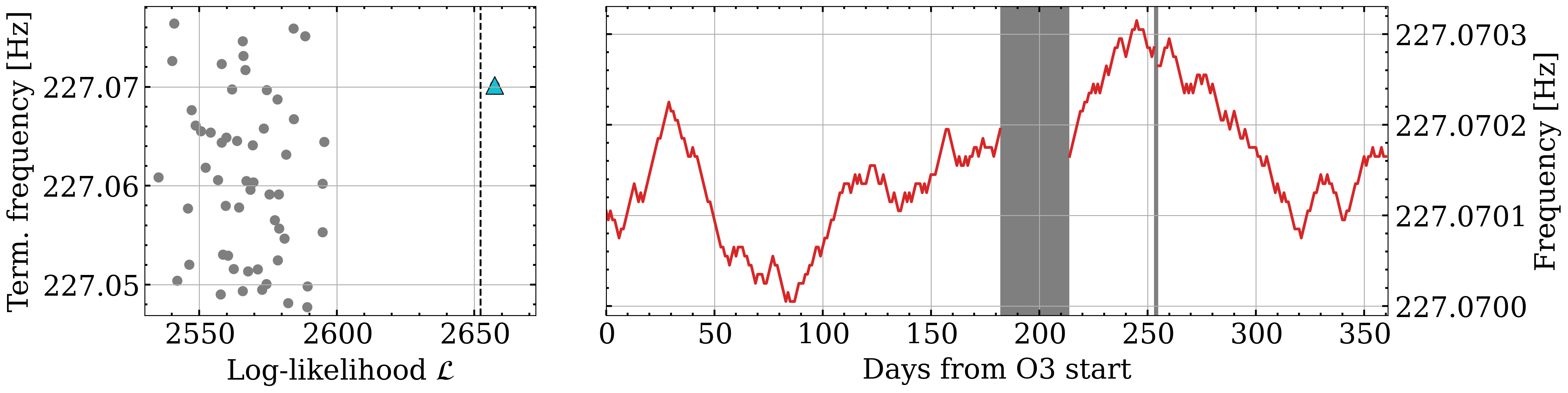}}
\hfill
\subfloat[\label{fig:outlier_d} Template: $\alpha=0.80$, $\delta=0.08$, $\dot{f}=-2\times 10^{-10}$ Hz/s.]{\includegraphics[width=1.3\columnwidth]{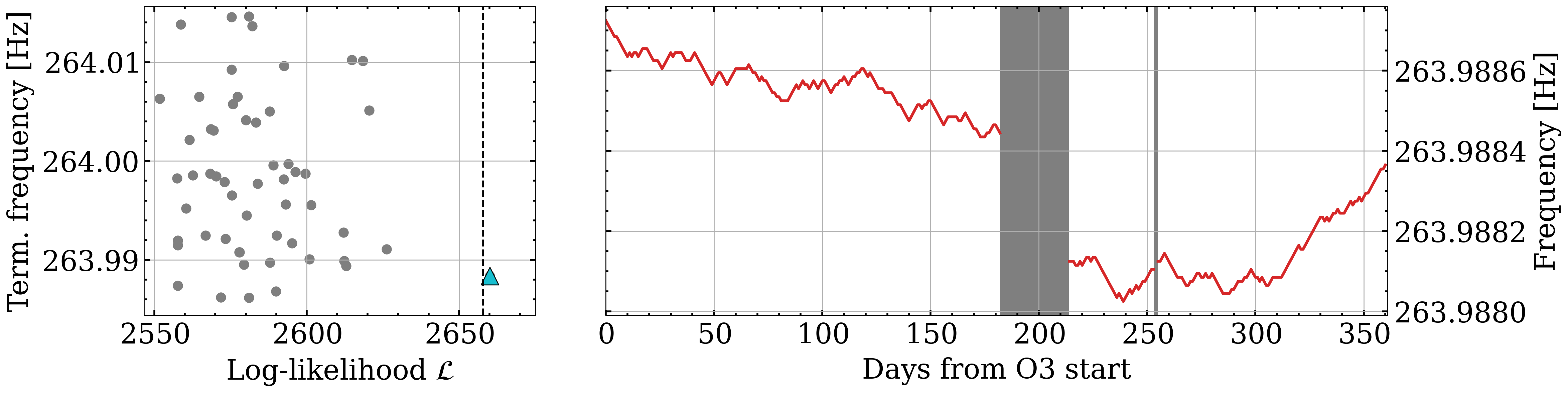}}
\hfill
\subfloat[\label{fig:outlier_e} Template: $\alpha=4.79$, $\delta=-0.70$, $\dot{f}=-6\times 10^{-10}$ Hz/s.]{\includegraphics[width=1.3\columnwidth]{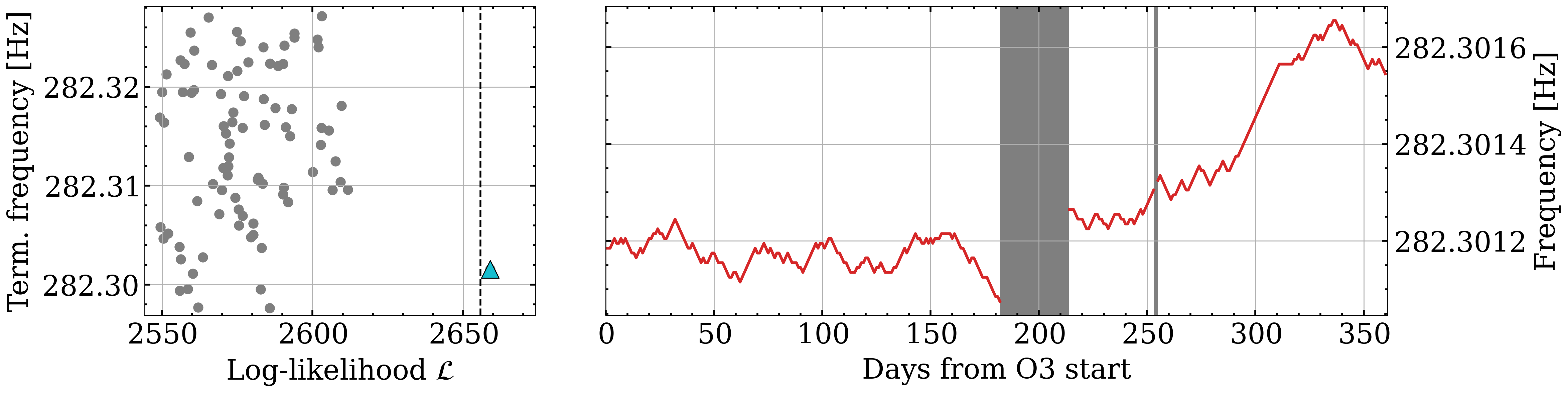}}
\caption{\label{fig:outlierfull} Detailed search results for the $15$ ASAF candidates which yielded an outlier. Left panels show the log-likelihood statistics for every searched template, with the threshold indicated by the vertical dashed line. Each template is associated with a frequency path terminating at a specific bin (after template demodulation), which is given on the vertical axis. Sub-threshold templates are shown by the gray points, and the above-threshold outlier is marked as a cyan triangle. Right panels show the optimal frequency path of the outlier template. The grey vertical bands indicate where no data is available; the large gap near the middle is the O3 commissioning break. The parameters of the outlier template, $(\alpha,\delta,\dot{f})$, are stated in the caption of each sub-figure, with sky positions in units of radians.}
\end{figure*}

\begin{figure*}
\ContinuedFloat
\subfloat[\label{fig:outlier_f} Template: $\alpha=3.89$, $\delta=-1.24$, $\dot{f}=0$ Hz/s.]{\includegraphics[width=1.3\columnwidth]{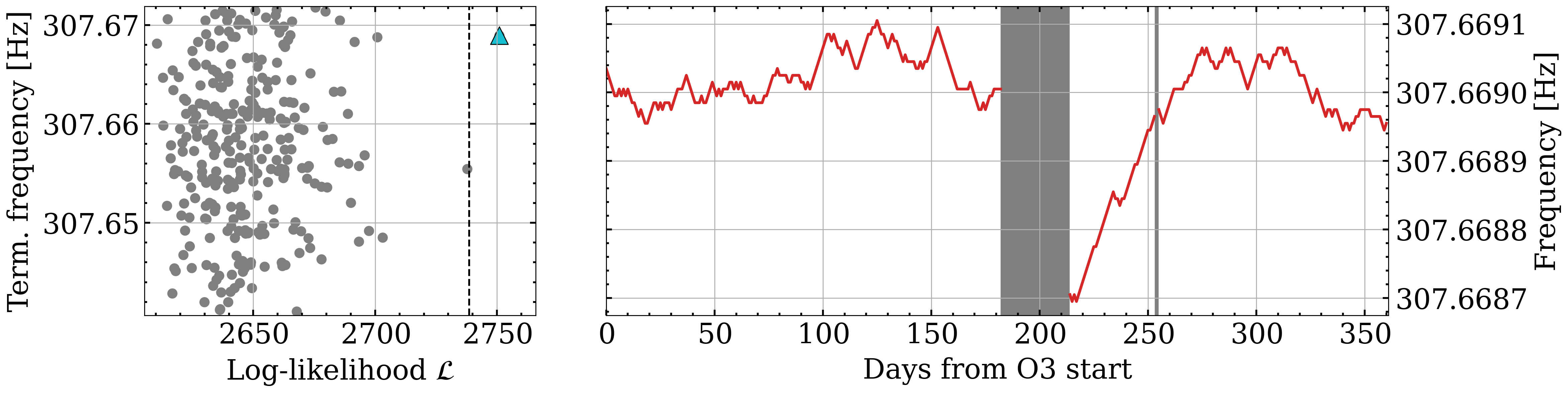}}
\hfill
\subfloat[\label{fig:outlier_g} Template: $\alpha=0.99$, $\delta=1.00$, $\dot{f}=6\times 10^{-10}$ Hz/s.]{\includegraphics[width=1.3\columnwidth]{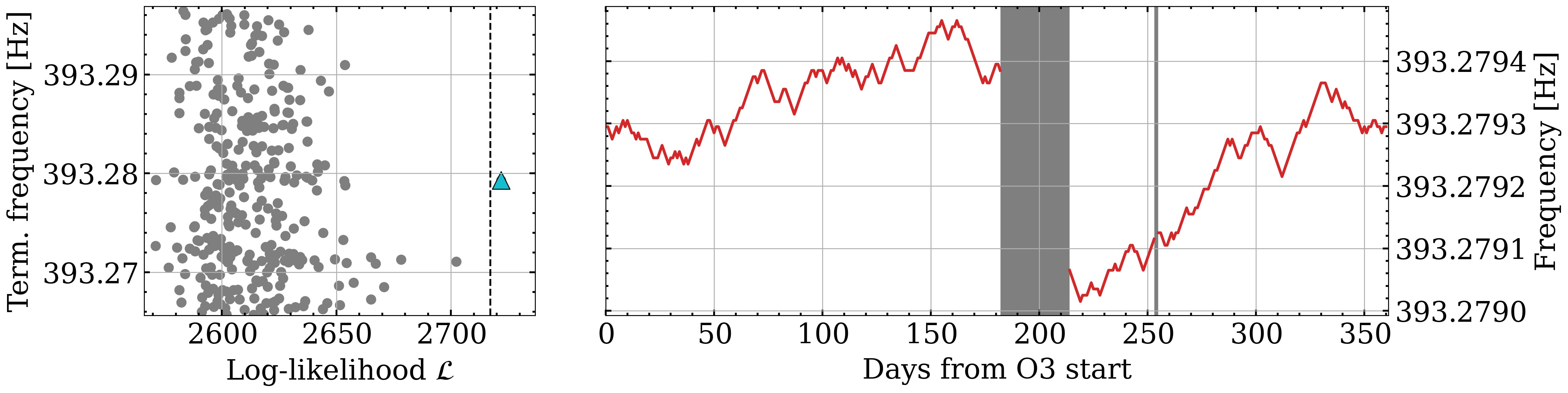}}
\hfill
\subfloat[\label{fig:outlier_h} Template: $\alpha=5.34$, $\delta=1.383$, $\dot{f}=-2\times 10^{-10}$ Hz/s.]{\includegraphics[width=1.3\columnwidth]{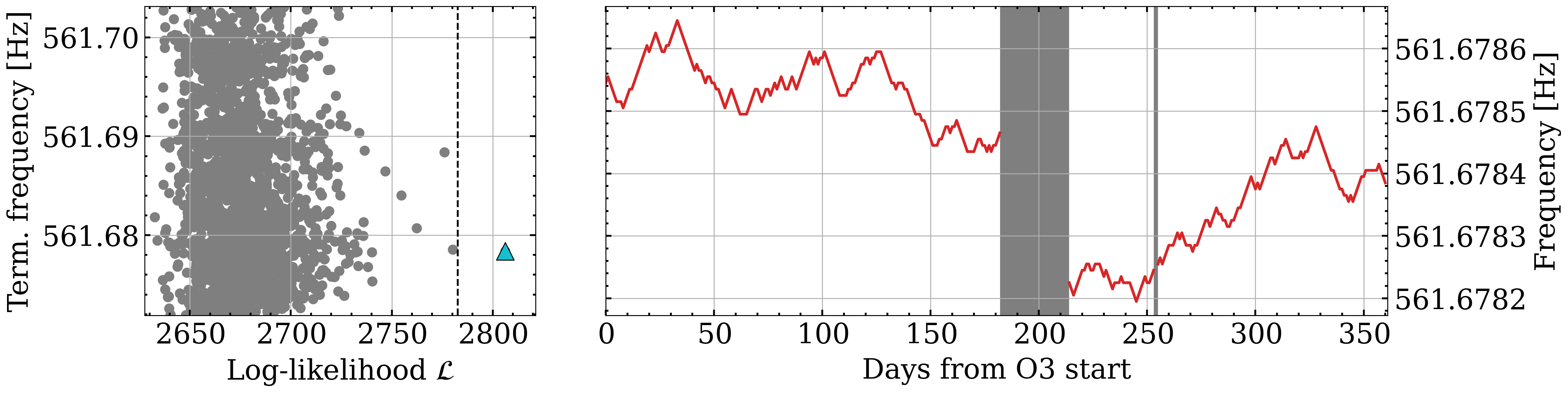}}
\hfill
\subfloat[Template: $\alpha=4.180$, $\delta=-0.46$, $\dot{f}=-6\times 10^{-10}$ Hz/s.]{\includegraphics[width=1.3\columnwidth]{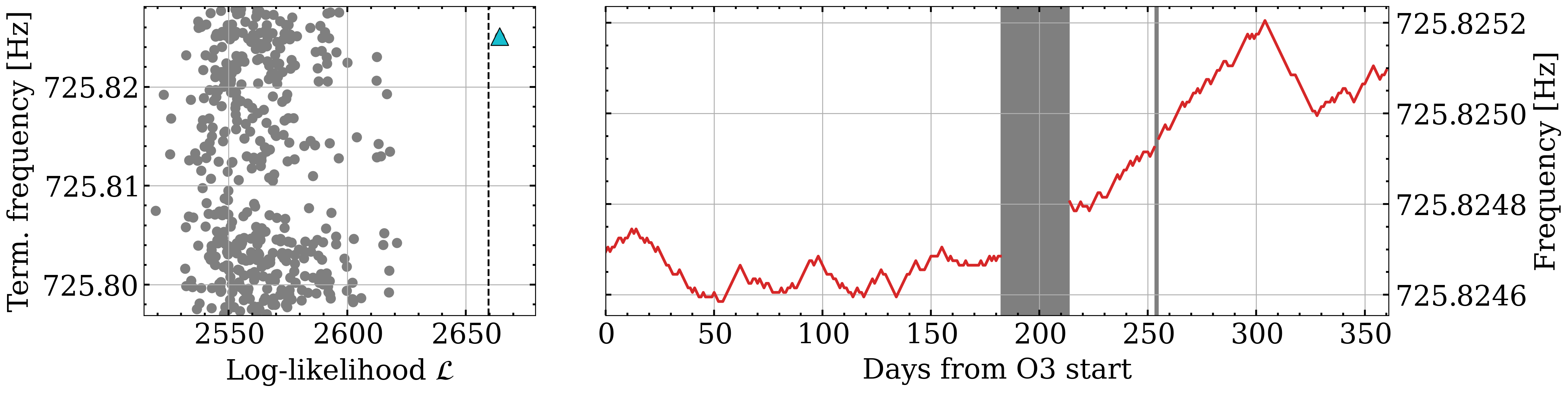}}
\hfill
\subfloat[\label{fig:outlier_j} Template: $\alpha=1.22$, $\delta=-1.339$, $\dot{f}=-1\times 10^{-9}$ Hz/s.]{\includegraphics[width=1.3\columnwidth]{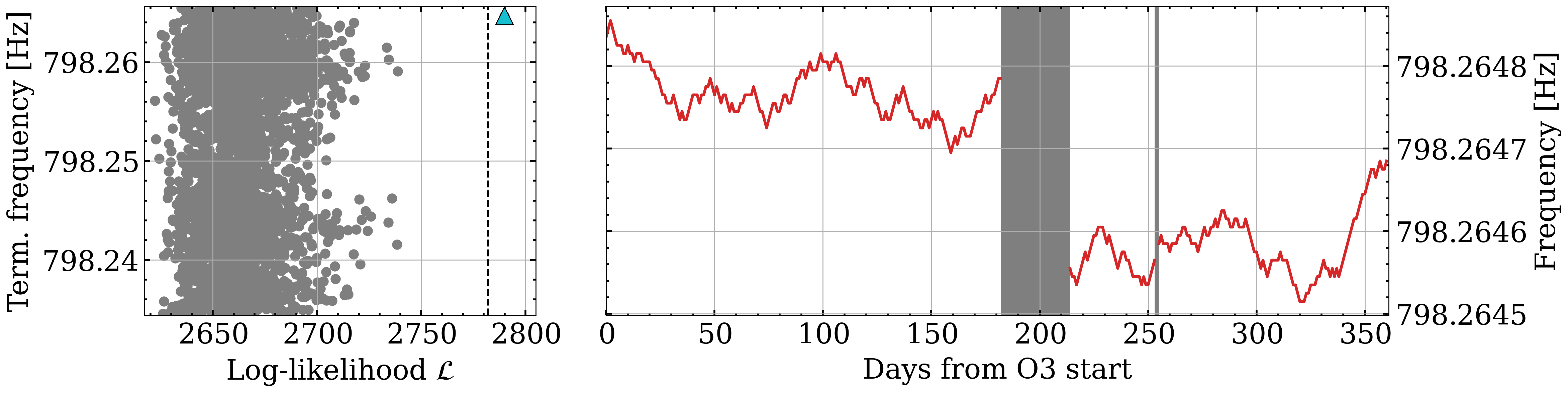}}
\caption{(cont.)}
\end{figure*}

\begin{figure*}
\ContinuedFloat
\subfloat[\label{fig:outlier_k} Template: $\alpha=3.77$, $\delta=1.091$, $\dot{f}=-1\times 10^{-9}$ Hz/s.]{\includegraphics[width=1.3\columnwidth]{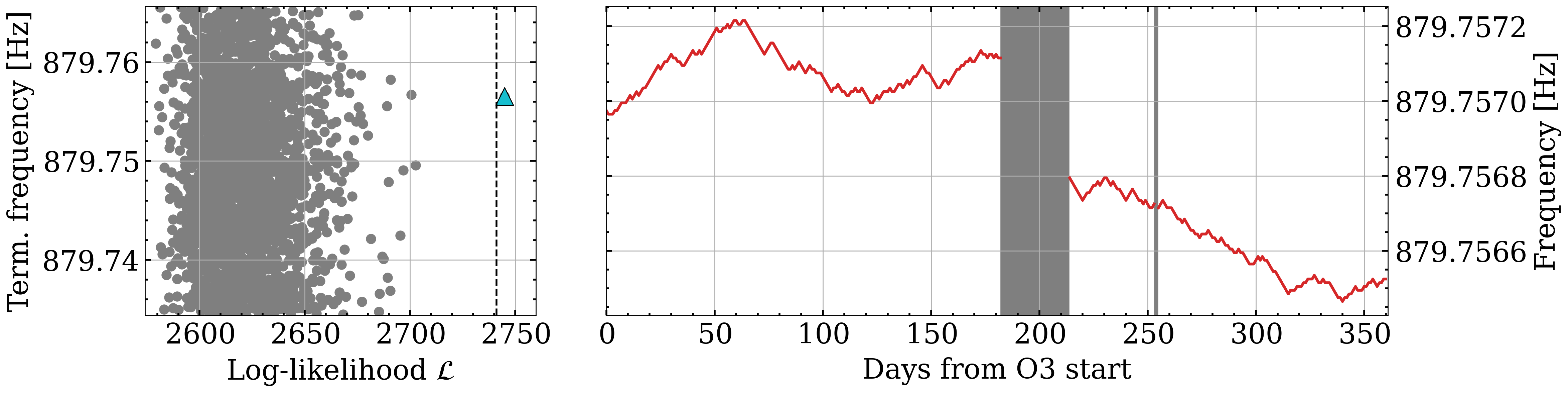}}
\hfill
\subfloat[\label{fig:outlier_l} Template: $\alpha=0.246$, $\delta=-0.582$, $\dot{f}=8\times 10^{-10}$ Hz/s.]{\includegraphics[width=1.3\columnwidth]{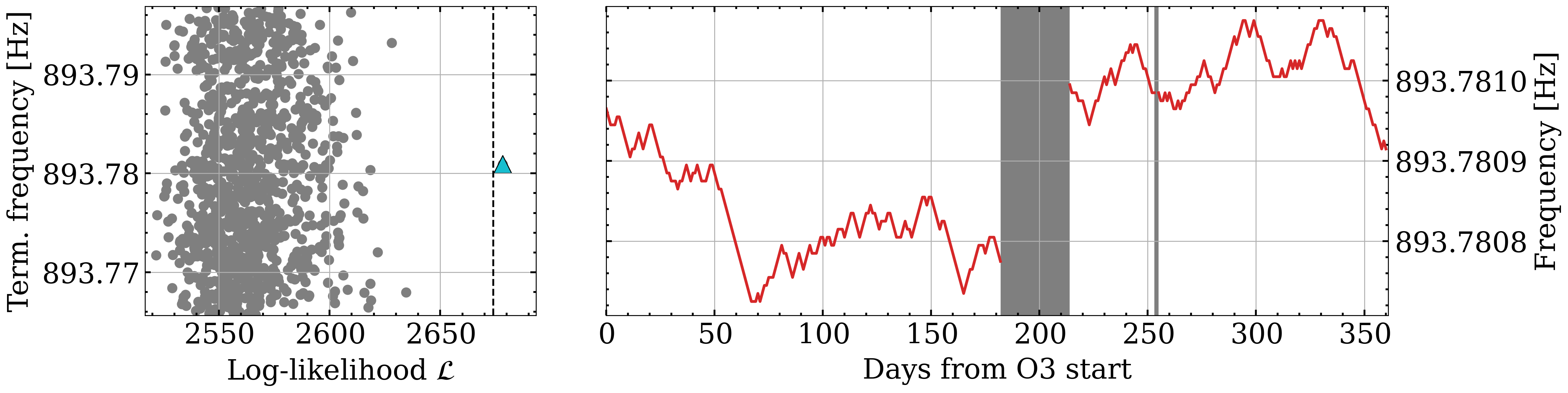}}
\hfill
\subfloat[\label{fig:outlier_m} (Vetoed) Template: $\alpha=4.546$, $\delta=0.537$, $\dot{f}=1\times 10^{-9}$ Hz/s.]{\includegraphics[width=1.3\columnwidth]{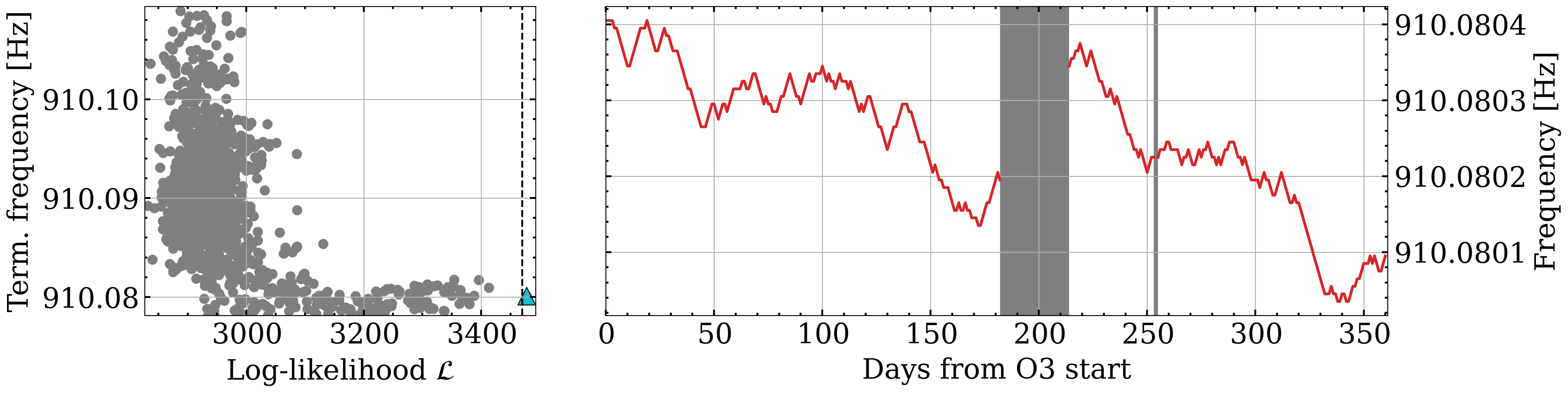}}
\hfill
\subfloat[\label{fig:outlier_n} Template: $\alpha=3.763$, $\delta=-0.533$, $\dot{f}=-2\times 10^{-10}$ Hz/s.]{\includegraphics[width=1.3\columnwidth]{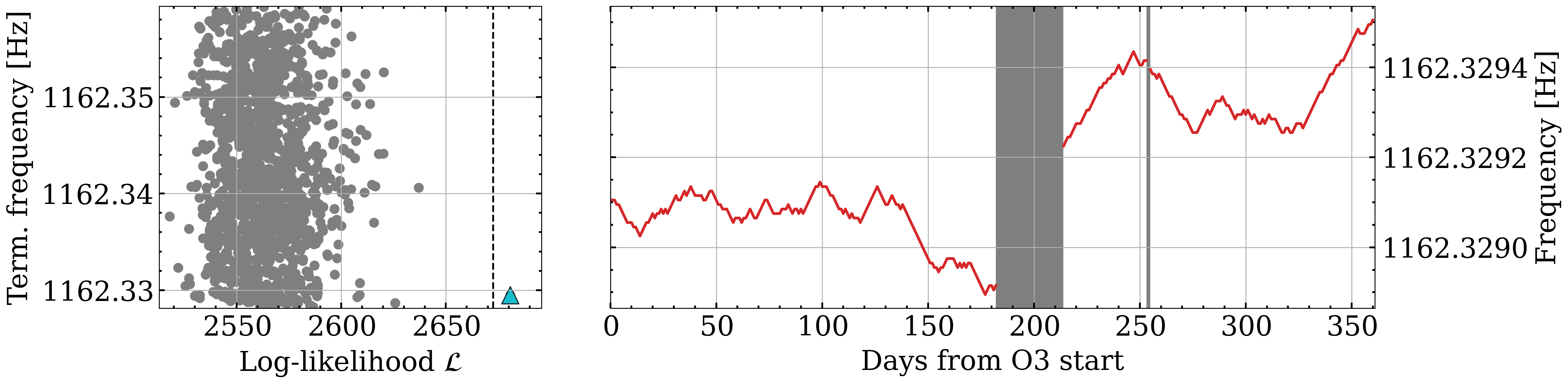}}
\hfill
\subfloat[\label{fig:outlier_o} Template: $\alpha=2.646$, $\delta=-0.563$, $\dot{f}=-2\times 10^{-10}$ Hz/s.]{\includegraphics[width=1.3\columnwidth]{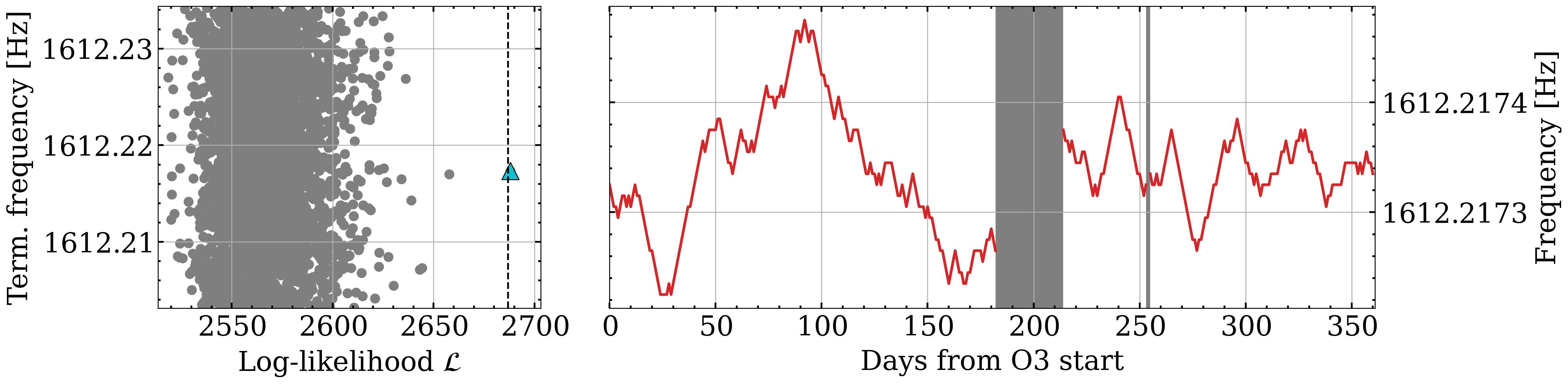}}
\caption{(cont.)}
\end{figure*}

\section{Number of outliers} \label{sec:nfa} 

Assuming all $14$ surviving outliers represent false alarms, we obtain fewer false alarms from our search than expected, given we targeted a $5\%$ false alarm probability per-candidate. There are a variety of possible causes for a lack of false alarms, such as a large fluctuation of the background, template correlations (which would result in an overestimate of the trials factor), or inaccuracy or bias in the fits to the off-target distributions. Regarding the fitting, we remind that reader that since we are following up ASAF candidates, we are in effect searching ``hotspots'' in the data, which may be biased towards containing unusual, non-Gaussian features. We could attempt calculating the thresholds on simulated Gaussian noise instead of using detector data, thereby avoiding complications from non-Gaussian data. Unfortunately, this is not ideal either since real detector noise is not perfectly stationary and Gaussian, and thresholds based on Gaussian noise would not reflect the true character of the data. As mentioned at the beginning of Sec.~\ref{sec:thresholds}, non-parametric methods for threshold estimation relax the assumption of an exponential tail, but are computationally unfeasible in a search such as ours.

Given that $14$ outliers is roughly half the mean expected number of false alarms, we might estimate that our $p_{\rm noise}$ values are a factor of two larger than they should be. Even if we halved $p_{\rm noise}$ for all outliers, none would be below a global false alarm probability corresponding to $5\%$ across all $515$ candidates, and therefore our earlier conclusion that no outliers are statistically significant remains unchanged.

\section{Correlation with ASAF SNR} \label{app:asaf}

\begin{figure}
\centering
\includegraphics[width=1\columnwidth]{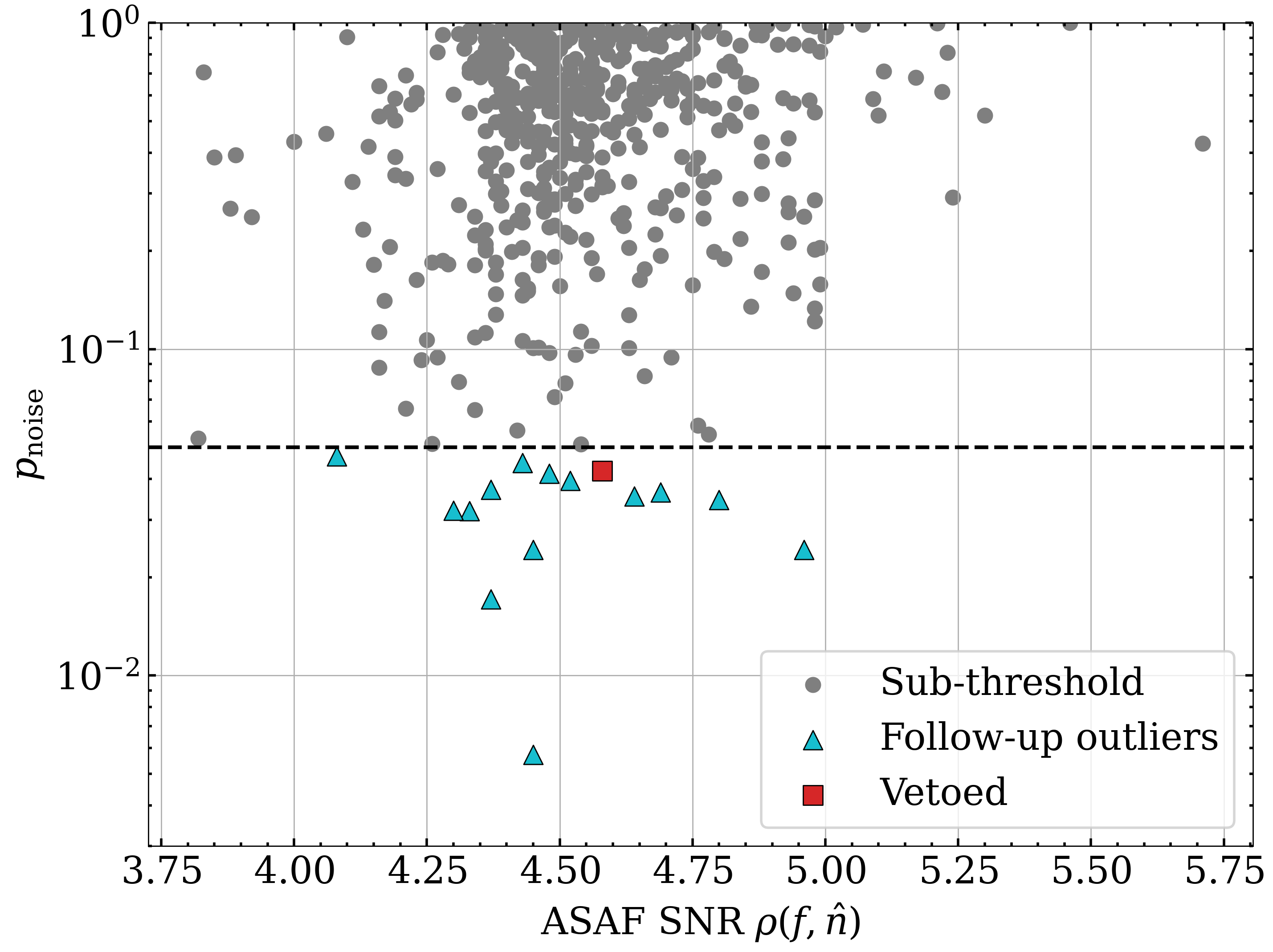}
\caption{\label{fig:asafviterbi} Relation between $p_{\rm noise}$ (vertical axis), defined in Eq.~\ref{eq:pnoise}, and the ASAF SNR (horizontal axis). Sub-threshold results are shown in grey, outliers are shown by the cyan triangles, and the red square is the vetoed outlier. The horizontal dashed line corresponds to a $5\%$ false alarm probability.}
\end{figure}

Fig.~\ref{fig:asafviterbi} shows $p_{\rm noise}$ plotted against the ASAF SNR statistic for each ASAF candidate. To test for any correlation, we compute the Pearson and Spearman's correlation coefficients between $p_{\rm noise}$ and ASAF SNR, and between $\mathcal{L}^*$ and ASAF SNR, for every ASAF candidate.  For the $\mathcal{L}^*$ and ASAF SNR comparison, the Pearson coefficient is $0.007$ with (two-sided) $p$-value $0.877$, and the Spearman's coefficient is $0.027$ with $p$-value $0.542$, indicating no evidence for a linear or monotonic correlation between the variates. For $p_{\rm noise}$ and ASAF SNR, the Pearson coefficient is $0.128$ with $p$-value $0.004$, and the Spearman's coefficient of $0.119$ with $p$-value $0.007$, suggesting that higher $p_{\rm noise}$ is positively correlated with higher ASAF SNR. This result is non-intuitive, as it suggests higher significance in the ASAF analysis is correlated with lower significance in our CW search, after accounting for trials factors in the CW search. Future studies may seek to investigate the correlation between CW and ASAF detection statistics when CW signals are injected upstream of both the ASAF and CW analyses.

\bibliography{main} 

\end{document}